\documentclass[final]{IEEEtran}
\usepackage[utf8]{inputenc}
\ifCLASSOPTIONcompsoc
    \usepackage[caption=false, font=normalsize, labelfont=sf, textfont=sf]{subfig}
\else
    \usepackage[caption=false, font=footnotesize]{subfig}
\fi
\usepackage{amsmath}
\usepackage{amsfonts}
\usepackage{booktabs}
\usepackage[noadjust]{cite}
\usepackage{tikz}
    \usetikzlibrary{arrows}
    \usetikzlibrary{shapes.geometric, positioning,quotes,chains} %Flowchart
    \usetikzlibrary{spy} %For inset zoom
    \usetikzlibrary{decorations.pathreplacing} %Curly bracket
    \usetikzlibrary{calc} %Relative coords
\usepackage{bm} % For bold vectors
\usepackage{xcolor}
\usepackage{pgfplots}
    \pgfplotsset{compat=1.14}
    \usepgfplotslibrary{patchplots} %FOR 3D Mesh
\usepackage{multirow}
\usepackage[draft]{fixme}
\usepackage{verbatim}
\usepackage{makecell}%Newline in table
\usepackage[binary-units=true]{siunitx}

\usepackage[noabbrev]{cleveref} %Should be last

\makeatletter %Arrow and tilde over letter
\newcommand{\pvec}{\@ifstar{\@pvecstar}{\@pvecnostar}}
\newcommand{\@pvecnostar}[1]{\,\widetilde{\!\bm{#1}}}
\newcommand{\@pvecstar}[2]{{\@pvecnostar{#1}}_{\mkern-1mu\relax#2}}
\makeatother
\crefname{figure}{Fig.}{Figs.}
\Crefname{figure}{Fig.}{Figs.}
\pgfplotscreateplotcyclelist{color_pattern}{%
draw={rgb:red,57;green,106;blue,177}\\%
draw={rgb:red,218;green,124;blue,48}\\%
draw={rgb:red,62;green,150;blue,81},thick,dash pattern=on 3pt off 1pt\\%
draw={rgb:red,204;green,37;blue,41},thick,densely dotted\\%
draw={rgb:red,83;green,81;blue,84},dashdotted, thick\\
draw={rgb:red,107;green,76;blue,154}\\
draw={rgb:red,146;green,36;blue,40},densely dashdotted,thick \\ draw={rgb:red,148;green,139;blue,61},dashed,thick\\
}

\title{An Expedient Approach to FDTD-based Modeling of Finite Periodic Structures}
\author{Aaron~J.~Kogon, and Costas~D.~Sarris,~\IEEEmembership{Senior Member, IEEE}}

\begin{document}
%\setkeys{Gin}{draft=false} %SHOW PICTURES IN DRAFT MODE
\maketitle

\begin{abstract}
This paper proposes an efficient FDTD technique for determining electromagnetic fields interacting with a finite-sized 2D and 3D periodic structures. The technique combines periodic boundary conditions---modelling fields away from the edges of the structure---with independent simulations of fields near the edges of the structure. It is shown that this algorithm efficiently determines the size of a periodic structure necessary for fields to converge to the infinitely-periodic case. Numerical validations of the technique illustrate the savings concomitant with the algorithm.
\end{abstract}

\section{Introduction}

Periodic structures are common geometries in electromagnetics, appearing in such forms as frequency selective surfaces, electromagnetic band gap media, photonic crystals, antenna arrays and metasurfaces, among others. 

A common way to simulate infinitely periodic structures in the Finite-Difference Time-Domain (FDTD) method \cite{taflove2005computational} is by employing periodic boundary conditions (PBCs) \cite{kogon2020fdtd}. PBCs collapse an infinitely-long periodic domain into a single unit cell. PBC simulations have been successfully used to extract transmission and reflection spectra \cite{yang2007simple}, attenuation constants \cite{kokkinos2006periodic, xu2007finite}, Brillouin diagrams \cite{kokkinos2006periodic, chan1995order}, among others.

Although PBCs are effective at simulating infinitely-periodic structures, the assumption of infinite extent is nonphysical. Fields obtained by simulations with PBCs are thus necessarily approximations of real finite-sized periodic structures. The approximations become increasingly worse closer to the edges of the structure, where the periodicity ends. Many analytical, semi-analytical and method-of-moments models for analyzing truncated structures have been discussed in the literature \cite{usoff1994edge, ishimaru1985finite,ko1988scattering,denison1995decomposition,wu1970analysis, capolino2000frequency}. In general, full-scale simulations of finite periodic structures---without PBCs---are typically run to precisely determine edge effects \cite{holter2002some}. Full-sized simulations can be computationally intensive. 

This paper advances a technique for accurately simulating fields interacting with finite-sized periodic structures. This hybrid method takes advantage of the periodicity of the structure while also effectively capturing the edge effects. 

First, the mathematical and theoretical background on periodic boundary conditions in FDTD is outlined. Next, the array scanning method (ASM), an algorithm used to remove unwanted periodic sources from PBC simulations, is presented. The methodologies for efficiently simulating finite two- and three-dimensional periodic structures are described and numerical results are provided. Convergence quantification metrics are defined to numerically measure accuracy of results obtained through the algorithm. 

Because of their computational efficiency, PBCs are heavily employed in the design of periodic structures. However, PBCs simulate infinite, practically unrealizable, structures. A pertinent question is how many unit cells of a finite periodic structure are needed to approximate the behavior of an infinite one. This question is addressed in the final section of this paper. 

\section{Periodic Boundary Conditions (PBCs)}

Before discussing the proposed algorithm, a mathematical background for FDTD PBCs is provided, with a discussion on the array scanning method (ASM) \cite{kogon2020fdtd}. 

Let \(\bm U(\bm r)\) represent an electromagnetic field in a three-dimensional periodic structure (periods \(d_x,d_y\) and \(d_z\) in the \(x,y\) and \(z\) directions respectively). Floquet's theorem states that for phasor \(\pvec{U}\):
\begin{align}
   \pvec{U}(\bm r + \bm d) = \pvec{U}(\bm r)e^{-j \bm k\cdot \bm d}\label{eq:floquet_phasor}
\end{align}
where \(\bm d = n_x d_x \bm{\hat x}+n_y d_y \bm{\hat y}+n_z d_z \bm{\hat z}\) is the lattice vector of the structure (\(n_x,n_y,n_z\in \mathbb Z\)) and where \(\bm{k} = k_{x}\bm{\hat x} + k_{y}\bm{\hat y} + k_{z}\bm{\hat z}\) is the reciprocal (Bloch) wave vector. 
Define the inverse Fourier transform of \(\pvec{U}\) as 
\begin{align}
    \bm U_s(\bm r,t) = \frac{1}{2\pi}\int_{-\infty}^\infty \pvec{U}(\bm r,\omega)e^{j\omega t}\,d\omega
\end{align}
so that the the real and imaginary parts of \(\bm U_s\) correspond to physical time-domain fields in quadrature. The field \(\bm U_s\) additionally satisfies the Floquet condition (\cref{eq:floquet_phasor}).

Both the real and imaginary components of \(\bm U_s\) may be simultaneously simulated using the FDTD algorithm. Then, the complex-valued Floquet condition, which relates fields one period apart, may be used as a periodic boundary condition over a unit cell. The wave vector \(\bm k\) of the real and imaginary parts of \(\bm U_s\) is set by the PBCs.

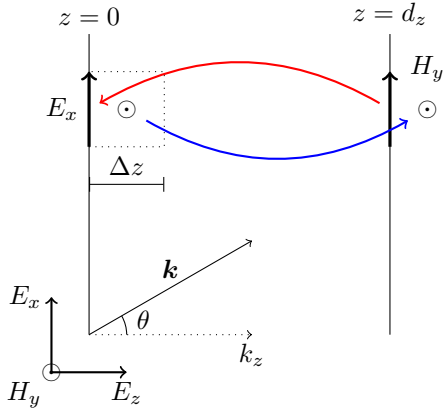
\begin{figure}
    \centering
    \begin{tikzpicture}
    \tikzset{myptr/.style={decoration={markings,mark=at position 1 with %
    {\arrow[scale=1,>=stealth]{>}}},postaction={decorate}}}
    
        \draw[dotted,->] (0,0) -- (2.5*0.866,0) node[below]{\(k_z\)};

        \draw[->] (0,0) -- (30:2.5) node[above,midway] {\(\bm{k}\)};
        
        \draw[] (0.5,0) arc (0:30:0.5);
        \node at ((15:.75) {\(\theta\)};

        \draw[] (0,0) -- (0,4) node[above] {\(z=0\)};
        
        \draw[] (4,0) -- (4,4) node[above] {\(z=d_z\)};

        \draw[dotted] (0,2.5) rectangle (1, 3.5);
        
        \draw[->,very thick] (0,2.5) -- (0,3.5) node[midway](ex_low){} node[left, midway] {\(E_x\)};
        \draw[->,very thick] (4,2.5) -- (4,3.5) node[midway](ex_high){};
        
        \draw node[](hy_low) at (.5, 3) {\(\odot\)};
        \draw node[label=above:\(H_y\)] (hy_high) at (4.5, 3) {\(\odot\)};
        
        \draw[red,thick,->] (ex_high) to[bend right] (ex_low) ;
        \draw[blue,thick, ->] (hy_low) to[bend right]  (hy_high);
        
        \draw[|-|] (0, 2) -- (1, 2) node[midway, above] {\(\Delta z\)};

        %%%
        \draw [->,thick] (-0.5, -0.5) -- (-0.5, 0.5) node[anchor=east] {\(E_x\)};
        \draw [->, thick] (-0.5, -0.5) node{\(\odot\)} node[anchor=north east]{\(H_y\)} -- (0.5, -0.5) node[anchor=north] {\(E_z\)};
    \end{tikzpicture}
    
    \caption{A period showing two-dimensional FDTD PBC updates (in the \(z\) direction). The \(E_x\) field at \(z=d_z\) is used to update the corresponding electric field at \(z=0\) (red arrow). The \(H_y\) field at \(z=\Delta z/2\) updates the \(H_y\) node at \(z=d_z+\Delta z/2\) (blue arrow).}
    \label{fig:updates_1D}
\end{figure}

To illustrate the implementation of FDTD PBCs, consider a structure that is periodic in the \(z\) direction with a unit cell\(d_z\), as shown in \cref{fig:updates_1D}. Introducing auxiliary \(H_y\) nodes one half-Yee cell outside the periodic domain at \(z = d_z+\Delta z/2\), the PBC update equations may be written as
\begin{align}
    E_x(x, 0) &= E_x(x, d_z)e^{j k_z d_z} \\
    H_y(x, d_z + \Delta z/2) &= H_y(x, \Delta z/2)e^{-j k_z d_z}
\end{align}
FDTD PBCs can impose periodicity in more dimensions similarly \cite{kogon2020fdtd}.

As mentioned, PBCs can be used to efficiently determine useful parameters and properties of periodic structures. They can be used to produce reflection and transmission spectra diagrams, Brillouin diagrams and determine attenuation constants. Of particular importance to this paper, however, is that PBCs can be used to determine fields due to finite sources interacting with periodic structures using the array scanning method.

\subsection{Array Scanning Method (ASM)}
\label{sec:asm}

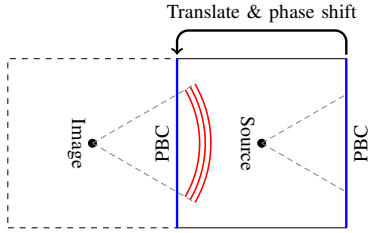
\begin{figure}
  \begin{center}
  \scalebox{0.75}{
    \begin{tikzpicture}
    \tikzset{
      pics/carc/.style args={#1:#2:#3}{
        code={
          \draw[pic actions] (#1:#3) arc(#1:#2:#3);
        }
      }
    }
    
    \draw (0,0) -- (3,0) -- (3,3) -- (0,3) -- cycle;    
    \draw[dashed] (0,0) -- (-3,0) -- (-3,3) -- (0,3) -- cycle;    
    
    \draw[thick, red] (-1.5,1.5) pic{carc=-30:30:1.9};
    \draw[thick, red] (-1.5,1.5) pic{carc=-30:30:2};
    \draw[thick, red] (-1.5,1.5) pic{carc=-30:30:2.1};
    
    \filldraw[black] (1.5,1.5) circle (2pt) node[rotate=270, anchor=north] {Source};
    \filldraw[black] (-1.5,1.5) circle (2pt) node[rotate=270, anchor=north] {Image};
    
    \draw[blue,very thick] (0,0) -- (0,3) node[black,anchor=south,midway,rotate=90] {PBC};
    \draw[blue,very thick] (3,0) -- (3,3) node[black,anchor=north,midway,rotate=90] {PBC};

    \draw[gray,densely dashed] (1.5,1.5) -- ++(30:{1.5/cos(30)});
    \draw[gray,densely dashed] (1.5,1.5) -- ++(-30:{1.5/cos(30)});

    \draw[gray,densely dashed] (-1.5,1.5) -- ++(30:{1.8/cos(30)});
    \draw[gray,densely dashed] (-1.5,1.5) -- ++(-30:{1.8/cos(30)});
    
    \draw[->, very thick,shorten >=2pt, shorten <= 2pt,  rounded corners=2mm] (3,3) -- (3,3.5) -- (0,3.5) node[midway,anchor=south]{Translate \& phase shift} -- (0,3);
    \end{tikzpicture}
    }
  \end{center}
  \caption{An illustration of a unit cell (right square) with PBCs (blue lines) on the left and right faces. A source in the unit cell produces fields (red) which travel rightwards within the dashed lines emanating from the source. The fields are then phase-shifted and translated to the left face, where they continue to propagate, as shown. These fields then appear as though they were generated by a phase-shifted image source in an adjacent unit cell (left square).}
  \label{fig:PBC_images}
\end{figure}

PBCs operate by translating fields from one edge of the periodic domain to the other side, with a complex phase shift. When fields produced by a source arrive at a PBC, they appear at the opposite side of the simulation with a phase shift (see \cref{fig:PBC_images}). This produces the effect of having infinite periods with sources in each period. The image sources have a linearly progressive phase shift from one unit cell to the next, set by the wave vector \(\bm k\) used in the PBC formulation. Explicitly, a simulation of a structure which is periodic in the \(z\) direction generates the field \(\bm U_\infty\) such that:
\begin{align}
    \bm U_\infty(x,y,t,k_y) = \sum_{n=-\infty}^\infty \bm {U}_{n}(x,y,t) e^{-jk_ynd_y}
\end{align}
where \(\bm {U}_{n}(x,y,t) = \bm {U}_{0}(x,y-nd_y,t)\) represents the fields generated by a source in unit cell \(n\).

The orthogonality of the complex exponential may be used to extract fields due to sources in unit cell \(n\):
\begin{align}
    \bm {U}_{n}(x, y, t) = \frac{d_y}{2\pi}\int_{-\pi/d_y}^{\pi/d_y} \bm {U}_{\infty}(x, y, t, k_y) e^{jk_y n d_y} \,dk_y. \label{eq:asm}
\end{align}
In three-dimensional periodic structure (size \(d_x\times d_y\times d_z\)), the field \(\bm{U}_{l,m,n}\) due to a source in the \((l,m,n)^\text{th}\) unit cell is:
\begin{align}
\begin{split}
    \bm {U}_{l,m,n}(\bm{r}, t) = &\frac{d_x d_y d_z}{(2\pi)^3} \int_{-\pi/d_x}^{\pi/d_x} \int_{-\pi/d_y}^{\pi/d_y} \int_{-\pi/d_z}^{\pi/d_z} \\
    &\bm {U}_{\infty}(\bm{r}, t, \bm{k}) e^{j \bm{k} \cdot \bm{d}} \,d^3\bm{k}\label{eq:asm_3d}
\end{split}
\end{align}
where \(\bm{d} = (ld_x, md_y, nd_z)\).

This integration procedure is known as the array scanning method \cite{munk1979plane,li2008efficient,capolino2007comparison}, which provides a technique for removing unwanted source images.

\Cref{eq:asm} may be approximated by numerical integration of fields determined by FDTD. \(\bm {U}_{\infty}\) can be evaluated from a single simulation, where \(k_y\) is set by the PBCs. The various simulations used to approximate \cref{eq:asm} are disjoint and may be run simultaneously.

One of the most robust quadrature techniques for estimating \cref{eq:asm} is the midpoint rectangular rule:
\begin{align}
    \begin{split}
        \bm U_n^M& = \frac{1}{M}\sum_{m=0}^{M-1} \bm I\left(k_y=\Delta k_y\left(m+\frac{1}{2}\right)-\frac{M}{2}\right)
    \end{split}
\end{align}
where \(\bm I=\bm {U}_{\infty}(x, y, t, k_y) e^{jk_y n d_y}\) is the integrand of \cref{eq:asm} and \(\Delta k_y = 2\pi/(M d_y)\). This formula removes the effect of image sources in \(M\) unit cells to either side of unit cell \(n\) exactly, up to a numerical error. The midpoint rectangular rule has been additionally found to outperform higher order integration methods \cite{capolino2007comparison}. 

\begin{figure}
  \begin{center}
  \scalebox{0.75}{
    \begin{tikzpicture}
    \tikzset{
      pics/carc/.style args={#1:#2:#3}{
        code={
          \draw[pic actions] (#1:#3) arc(#1:#2:#3);
        }
      }
    }
    
    \draw (0,0) -- (1,0) -- (1,2) -- (0,2) -- cycle;
    \draw (1,0) -- (2,0) -- (2,2) -- (1,2) -- cycle;
    
    \node[] at (2.5,1) {\(\cdots\)};
    \draw (4,0) -- (5,0) -- (5,2) -- (4,2) -- cycle;
    \draw (3,0) -- (4,0) -- (4,2) -- (3,2) -- cycle;
    
    \filldraw[black] (0.75,1) circle (2pt) node[rotate=270, anchor=north] {Source};
    
    \filldraw[black] (4.75,1) circle (2pt) node[rotate=270, anchor=north] {Image};
    
    \draw[thick, dashed] (1.25, 0) -- (1.25, 2);
    \draw[thick, dashed] (0.25, 0) -- (0.25, 2);
    
    \draw[|-|] (0.75,2.15) -- (1.25,2.15) node[anchor=south,midway]{\(a\)};
    
    \draw[|-|] (0, -0.15) -- (1, -0.15) node[anchor=north,midway]{\(d\)};
    
    \draw[|-|] (1.25, -0.15) -- (4.75, -0.15) node[anchor=north,midway]{\(Md-a\)};
    
    \draw[thick, red] (0.75,1) pic{carc=-15:15:3.2};
    \draw[thick, red] (0.75,1) pic{carc=-15:15:3.3};
    \draw[thick, red] (0.75,1) pic{carc=-15:15:3.4};
    
    \draw[thick, red] (4.75,1) pic{carc=165:195:3.2};
    \draw[thick, red] (4.75,1) pic{carc=165:195:3.3};
    \draw[thick, red] (4.75,1) pic{carc=165:195:3.4};
    
    \end{tikzpicture}
    }
  \end{center}
  \caption{An illustration of the ASM method removing unwanted image sources in a periodic structure (period \(d\)). The ASM integration order \(M\) determines the location of the nearest parasitic image source. By selecting \(M\) to be sufficiently large, fields (shown in red) generated by the image do not enter a region of interest \(a\) units to either side of the source within runtime. By the end of the simulation, the wavefronts from unwanted sources travel a distance of at most \(Md-a\).}
  \label{fig:ASM_schematic}
\end{figure}
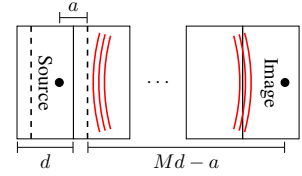

Using the midpoint rectangular rule, it is straightforward to select quadrature order \(M\) to be large enough that fields from parasitic images do not enter a region of interest within a time \(t_0\). Suppose fields in a region \(a\) units to either side of a source are desired in a periodic structure of period \(d\) (shown in \cref{fig:ASM_schematic}). If \(M\) is sufficiently large, the wavefront from a parasitic image source will reach the boundary of the region of interest at the end of the simulation, meaning \(ct_0 = Md-a\) where \(c\) is the speed of light. Selecting the integration order to be at least \(M = \lceil (t_0c+a)/d \rceil\) ensures that fields from unwanted image sources will not enter the region of interest by \(t_0\).

Symmetry with respect to the periodic directions (as in \cref{fig:2d_edges}) may be exploited to reduce the ASM integration order. When the field generated by sources in the center period \(\bm {U}_{0}\) is even:
\begin{align}
\begin{split}
    \bm {U}_{n}(x, y, t) = 
    \frac{d_y}{2\pi}&\int_{0}^{\pi/d_y} \left( \bm {U}_{\infty}(x, y, t, k_y) e^{jk_y n d_y} + \right.\\
    &\left. \bm {U}_{\infty}(x, -y, t, k_y) e^{-jk_y n d_y}\right)\,dk_y .
\end{split}
\end{align}
When \(\bm {U}_{0}\) is odd:
\begin{align}
\begin{split}
    \bm {U}_{n}(x, y, t) = 
    \frac{d_y}{2\pi}&\int_{0}^{\pi/d_y} \left( \bm {U}_{\infty}(x, y, t, k_y) e^{jk_y n d_y} - \right.\\
    &\left. \bm {U}_{\infty}(x, -y, t, k_y) e^{-jk_y n d_y}\right)\,dk_y .
\end{split}
\end{align}
The domain of integration of these integrals is half of that in \cref{eq:asm}. Therefore, the computation of these integrals requires half as many simulations to be approximated at a given precision.

\section{General Methodology}\label{sec:methodology}

Simulations with PBCs produce fields which behave as though they interact with an infinitely periodic structure. Therefore, fields obtained through simulations with PBCs may be reasonably used as approximations of fields interacting with finite periodic structures far from the edges of the structure where the periodicity ends.

\begin{figure}
    \centering
    \def \period {0.4}
    \def \unit cell {0.4}
    \def \gwidth {0.3}
    \def \widthdiff {0.5*\period-0.5*\gwidth}
    \begin{tikzpicture}
                
        \foreach \x in {-3,...,3}
            \draw[fill=white!50!gray] (\period*\x - 0.5*\gwidth,0.05) rectangle (\period*\x+\gwidth - 0.5*\gwidth, -0.05);

        \foreach \x in {-6,...,6}
            \draw[fill=white!50!gray] (\period*\x - 0.5*\gwidth,0.05+2) rectangle (\period*\x+\gwidth - 0.5*\gwidth, -0.05 +2);
        %\draw[fill] (8*\period, 0.5) circle (2pt) node[above=2pt] {Source};
        
        \foreach \x in {-3,...,3}
            \draw[fill=white!50!gray] (\period*\x - 0.5*\gwidth,0.05-3) rectangle (\period*\x+\gwidth - 0.5*\gwidth, -0.05-3);
            
        \draw[fill=white,draw=white] (-1.0*\period - 0.5*\gwidth,0.1) rectangle (1.0*\period+\gwidth - 0.5*\gwidth, -0.1);
        
        \draw[red] plot[samples=200,domain=-6*\period:6*\period,smooth, variable=\x] ({\x}, {exp(-\x*\x)+2.5});
        \draw[green, very thick] plot[samples=200,domain=-2*\period:2*\period,smooth, variable=\x] ({\x}, {exp(-\x*\x)+2.5});
        
        \node at (10*\period, 2.5) {Infinite structure};
        \node at (7*\period, 2) {\textbf{...}};
        \node at (-7*\period, 2) {\textbf{...}};

        \draw[green, very thick] plot[samples=200,domain=-6*\period:-2*\period,smooth, variable=\x] ({\x}, {exp(-\x*\x) + cos(610*\x)*\x*\x/20+0.5});

        \draw[red] plot[samples=200,domain=-2*\period:2*\period,smooth, variable=\x]({\x}, {exp(-\x*\x) + cos(610*\x)*\x*\x/20+0.5});
        \draw[green, very thick] plot[samples=200,domain=2*\period:6*\period,smooth, variable=\x] ({\x}, {exp(-\x*\x) + cos(610*\x)*\x*\x/20+0.5});
        \node at (10*\period, 0.5) {Edges};
        
        \draw[green, very thick] plot[samples=200,domain=-6*\period:-2*\period] ({\x}, {exp(-\x*\x) + cos(610*\x)*\x*\x/20-2.5});
        \draw[green, very thick] plot[samples=200,domain=-2*\period:2*\period] ({\x}, {exp(-\x*\x) -2.5});
        \draw[green, very thick] plot[samples=200,domain=2*\period:6*\period] ({\x}, {exp(-\x*\x) + cos(610*\x)*\x*\x/20-2.5});
        
        \draw[gray, very thick, ->] (0, -0.5) -- (0, -1.25);
        
        \node[] at (10*\period,  -2.5) {Merged fields};
        
        \draw[densely dotted, blue, very thick] (-2*\period, -3) -- (-2*\period, 3.5);
        \draw[densely dotted, blue, very thick] (2*\period, -3) -- (2*\period, 3.5);
        
        %\draw[<->, shorten >=2pt, shorten <=2pt] (4.5*\period, -0.3) -- (11.5*\period, -0.3) node[midway,below] {\(2N^I+1\) cells};
        
        %\draw[<->, shorten >=2pt, shorten <=2pt] (0.5*\gwidth, 0.3) -- (4.5*\period, 0.3) node[midway,above] {\(N^E\) cells};
        %\draw[<->, shorten >=2pt, shorten <=2pt] (11.5*\period, 0.3) -- (16*\period-0.5*\gwidth, 0.3) node[midway,above] {\(N^E\) cells};
        %%
        
        \node[align=center] at (-4.5*\period, 4.5) {Edge\\region};
        \node[align=center] at (0, 4.5) {Inner\\region};
        \node[align=center] at (4.5*\period, 4.5) {Edge\\region};
        
    \end{tikzpicture}
    \caption{The fields in the inner region (between the vertical blue lines) of a finite structure can be estimated by fields from an infinite structure (top). The fields along the edges of the structure (outside the blue lines) can be calculated by simulating the edges of the finite structure (middle). The fields can be fused together to create a continuous field approximating those interacting with the finite structure (bottom).}
    \label{fig:merge}
\end{figure}
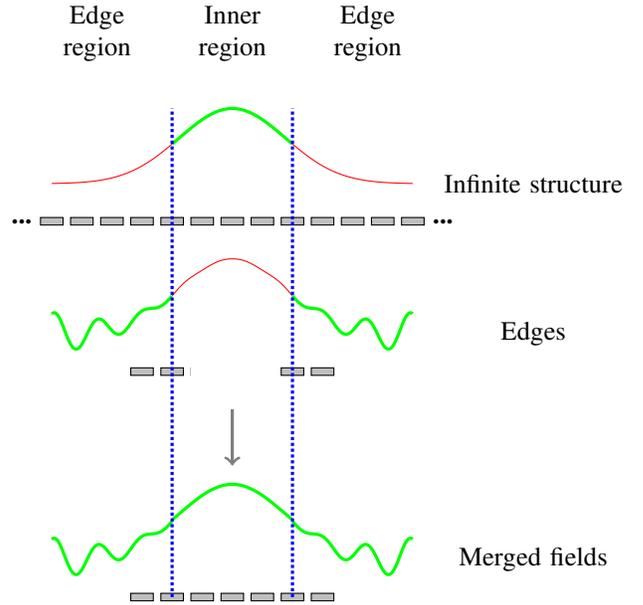

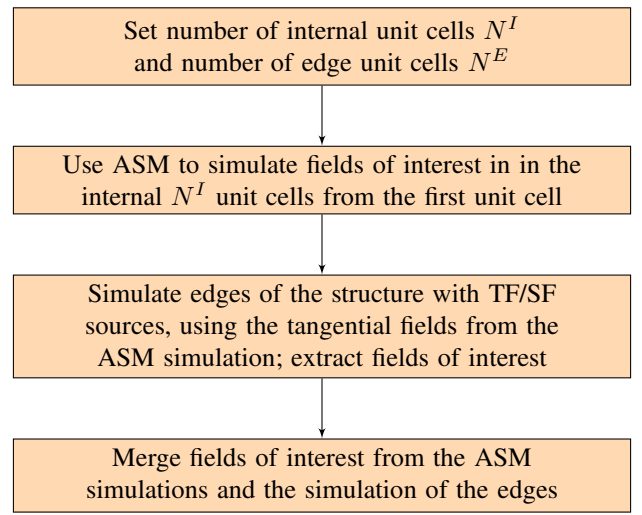
\begin{figure}
    \centering
    %%%%%    
    \begin{tikzpicture}[
    line/.style = {draw, -latex'},
    node distance = 8mm and 16mm,
      start chain = A going below,
      base/.style = {draw, minimum width=32mm, minimum height=8mm,
                     align=center, on chain=A},
        cloud/.style = {base, draw, ellipse,fill=green!20,},
 startstop/.style = {base, rectangle, rounded corners, fill=red!30},
   process/.style = {base, rectangle, fill=orange!30},
        io/.style = {base, trapezium, 
                     trapezium left angle=70, trapezium right angle=110,
                     fill=blue!30},
  decision/.style = {base, diamond, fill=green!30},
  every edge quotes/.style = {auto=right}]
        % Place nodes
        %\node [startstop] (start)      {Start};
        \node [process,text width=0.9\columnwidth] (setag) {Set number of internal unit cells \(N^I\) and number of edge unit cells \(N^E\)};
        \node [process,text width=0.9\columnwidth] (asm)        {Use ASM to simulate fields of interest in in the internal \(N^I\) unit cells from the first unit cell};
        \node [process,text width=0.9\columnwidth] (edges)         {Simulate edges of the structure with TF/SF sources, using the tangential fields from the ASM simulation; extract fields of interest };
        \node [process,text width=0.9\columnwidth]  (merge)       {Merge fields of interest from the ASM simulations and the simulation of the edges};
        %\node [decision] (converged)       {Converged?};
        %\node [startstop]  (end) {Stop};
        %\node [process, right=of converged]    {Increase \(G\)};

%%%%%%%%%%%%%%%
        %\path [line] (start) -- (setag);
        \path [line] (setag) -- (asm);
        \path [line] (asm) -- (edges);
        \path [line] (edges) -- (merge);
        %\path [line] (merge) -- (end);
    \end{tikzpicture}
    \caption{A flowchart describing the procedure of efficiently determining fields interacting with finite periodic structures and estimating the accuracy of the result. Note that the ASM step may be used to determine tangential fields along the boundaries of many differently-sized regions simultaneously. Subsequently, variously-sized edge-unit cell simulations can be run simultaneously as well.}
    \label{fig:edges_flowchart}
\end{figure}

Fields determined by PBC simulations may then be injected into a full-size FDTD simulation of the edge unit cells using the Total-Field Scattered-Field (TF/SF) source condition \cite{taflove2005computational}. With the TF/SF boundary ``replaying'' the fields from the PBC simulation, only the edges of the structure need to be simulated to sufficiently capture the edge effects. When there is a sufficient number of periods included in the simulation of the edges of the structure, there is a smooth transition between fields obtained by the ASM and fields obtained from the simulation of the edges. At this point, fields obtained by the ASM and fields obtained from the simulation of the edges may be patched together to obtain a continuous field. \Cref{fig:merge} shows how the infinite structure provides an estimate for the fields in the inner region (far from the edges), and a simulation of the edges alone can estimate the edge fields. The fields of the inner and edge regions are pasted together at the interfaces between the regions, as shown. The procedure is outlined schematically in \cref{fig:edges_flowchart}.

The TF/SF boundary reproduces the fields propagating along an infinitely periodic structure. If there were an infinite number of edge unit cells, there would be no fields in the scattered-field (SF) region. Since the simulation of the edges is finite, however, some fields will not behave precisely as the TF/SF boundary anticipates, so some fields will appear behind the boundary. The fields in the SF region become less pronounced as more periods along the edge are simulated. In any case, an absorber should be placed behind the TF/SF boundary in the SF region to absorb these fields and prevent them from being reflected back.

Some time \(t_s\) is required for wavefronts to reach the planes along which sampling takes place in the ASM simulations. This means that time \(t_s\) is required to elapse before the TF/SF boundaries start emitting any fields at all. For instance, consider a source positioned in a periodic structure as \cref{fig:2d_edges}. If \(l\) is the shortest distance between the source and a one of the sides of the inner region (where sampling takes place), waves arrive arrive at the nearest inner region boundary at time \(t_s=l/c\) or later. Thus, there is no need to record (or replay) fields before time \(t_s\).

%Note that uniform structures, such as slabs, may be modelled as periodic structures with an arbitrary period. In this case, the choice of period size affects only the ASM simulations---selecting a smaller period increases ASM integration order while decreasing the runtime of each ASM simulation; selecting a larger period has the opposite effect.

%Patching fields from simulations from the ASM and from the independent simulation of the edges can be done simply by pasting fields at the interface (see \Cref{fig:merge}). In many cases, when the transition is sufficiently smooth, the two fields may simply be fused together, as is done in this paper.

It should be noted that only one set of ASM simulations is necessary to extract tangential fields over internal regions of several sizes. This is possible because the ASM may be used to extract fields anywhere along the infinitely-periodic structure, given that the ASM integration order is sufficiently high. Additionally, several simultaneous simulations of the edges may be carried out by parallelization. 

In a two-dimensional simulation, a periodic surface has two disconnected edges (see \cref{fig:2d_edges}), in which case each edge may be simulated independently. A three-dimensional periodic structure has edges continuously running along the edges of a rectangle, shown in \cref{fig:3d_edges}, which inhibits the separation of the edges from each other. The two- and three-dimensional cases are discussed individually below.

%Noteworthy is the fact that increasing the size of a finite periodic structure by a factor \(\alpha\) in each of \(p\) periodic directions increases the computational resources (memory consumption, number of updates) as \(O(\alpha^p)\) in a full-sized simulation. However, if one were to likewise scale size of a simulation of \(N^E\) edge cells (in each periodic direction) in this fashion, the computational resources would scale as \(O(\alpha)\).

In sum, the proposed algorithm may be broken into a two-step process. First, the ASM is used to record fields tangential to the periodic faces (several periods away from the source) as a function of time. Second, a simulation of the edges is performed, where the fields previously determined by the ASM are injected into the simulation using the TF/SF source formulation.

All simulations below were performed using MATLAB on a computer with a quad-core \SI{3.00}{\giga\hertz} Intel Core i5-7400 processor and \SI{8}{\giga\byte} of RAM. 

\section{2D Finite-Sized Structure Simulations}
\label{sec:2d_edge_effects}
Two-dimensional simulations may be used to model structures and sources with infinite length in one axis. In many situations, two-dimensional simulations can be used to efficiently approximate physical structures.

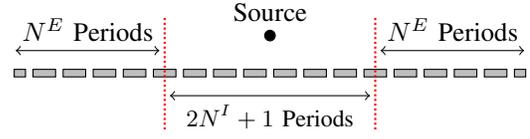
\begin{figure}
    \centering
    \def \period {0.4}
    \def \gwidth {0.3}
    \def \widthdiff {0.5*\period-0.5*\gwidth}
    \begin{tikzpicture}
        \foreach \x in {-1,...,16}
            \ifthenelse{\x=-1 \OR \x=16}
                {
                    \ifthenelse{\x=-1}
                    {\draw[fill=white!50!gray] (0.5*\gwidth+\period*\x +\widthdiff,0.05) rectangle (\period*\x+\gwidth + \widthdiff, -0.05);}
                    {\draw[fill=white!50!gray] (\period*\x +\widthdiff,0.05) rectangle (\period*\x+0.5*\gwidth + \widthdiff, -0.05);}
                }
                {\draw[fill=white!50!gray] (\period*\x +\widthdiff,0.05) rectangle (\period*\x+\gwidth + \widthdiff, -0.05);};
            
        \draw[fill] (8*\period, 0.5) circle (2pt) node[above=2pt] {Source};

        \draw[densely dotted, red, thick] (4.5*\period, -0.75) -- (4.5*\period, 0.75);
        \draw[densely dotted, red, thick] (11.5*\period, -0.75) -- (11.5*\period, 0.75);
        
        \draw[<->, shorten >=2pt, shorten <=2pt] (4.5*\period, -0.3) -- (11.5*\period, -0.3) node[midway,below] {\small \(2N^I+1\) Periods};
        
        \draw[<->, shorten >=2pt, shorten <=2pt] (-1*\period+0.5*\gwidth, 0.3) -- (4.5*\period, 0.3) node[midway,above] {\(N^E\) Periods};
        \draw[<->, shorten >=2pt, shorten <=2pt] (11.5*\period, 0.3) -- (17*\period-0.5*\gwidth, 0.3) node[midway,above] {\(N^E\) Periods};

    \end{tikzpicture}
    \caption{A periodic grating in two dimensions. The inner region of \(2N^I+1\) unit cells may be approximated by the ASM, and the outer edge regions of \(N^E\) unit cells may be simulated separately.}
    \label{fig:2d_edges}
\end{figure}

Consider a structure made up of \(N^E\) edge unit cells at either side of the structure and \(2N^I+1\) internal unit cells as shown in \cref{fig:2d_edges}. First, a single unit cell of the structure is simulated with PBCs (as in \cref{fig:grating_pbc}). The ASM is used to determine electric and magnetic fields perpendicular to the direction of periodicity, \(N^I\) periodic unit cells to either side of the first unit cell as functions of time. The position on the computational unit cell is shown as a red dotted line in \cref{fig:grating_pbc}. That is, the ASM may be used to record the fields
\begin{alignat}{2}
   &E_z(x,(N^I+1/2)d_y,t), &&H_x(x,(N^I+1/2)d_y,t)\\
    \intertext{and}
    &E_z(x,-(N^I+1/2)d_y,t), &&\quad H_x(x,-(N^I+1/2)d_y,t).
\end{alignat}

The recorded fields can then be ``replayed'' in a simulation of a finite number of unit cells to model the edge of periodic structure (as in \cref{fig:grating_edge}).  Some space in the SF region may be used as buffer between the source and the PML.

\subsection{Finite-Sized 2D Subwavelength Grating}\label{sec:subwavelength_grating}

In order to illustrate the procedure, a 2D transverse-electric (TE) simulation of a 200-unit cell subwavelength metallic grating was performed. Subwavelength gratings have been studied for their ability to convert evanescent waves into propagating waves, making them useful in lensing applications \cite{memarian2012evanescent, liu2007far}. They have also been used as reflectors \cite{mateus2004ultrabroadband} and optical switches \cite{min2008all}.

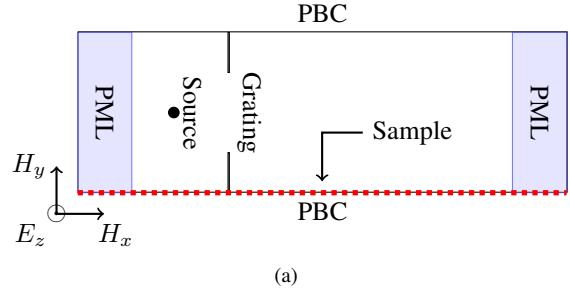
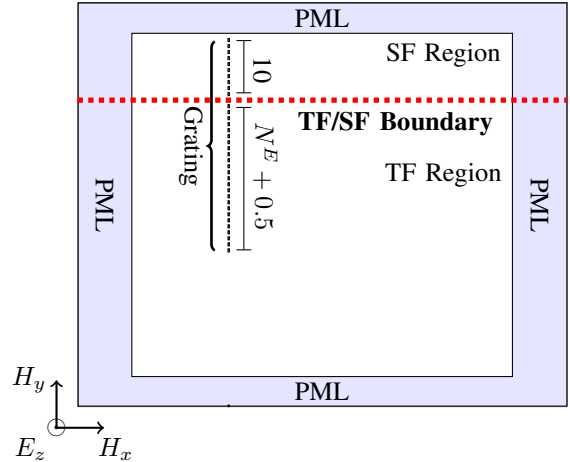
\begin{figure}
    \def\dx{0.018cm}\def\dy{0.26668cm} 
    \centering
    \begin{minipage}{\columnwidth}
    \subfloat[]{
    \begin{tikzpicture}[x = \dx, y = \dy] % Measure in cells!
        %\draw (0,0) rectangle (120, 32);
        
        \node[draw,anchor=south west,  minimum width=361*\dx, minimum height=8*\dy, label=below:PBC, label=above:PBC] (bound) at (0,0) {};
        
        \filldraw[fill=blue!20,draw=blue,opacity=0.5] (0,0) rectangle (40, 8) node[pos=.5, rotate=270,opacity=1] {PML};
        
        \filldraw[fill=blue!20,draw=blue,opacity=0.5] (361,0) rectangle (361-40, 8) node[pos=.5, rotate=270,opacity=1] {PML};
        
        %\filldraw[fill=black!20,draw=black] (50,8) rectangle (51, 24) node[pos=.5, rotate=90,anchor=north] {Grating};
        %\filldraw[fill=black!20,draw=black] (50,0) rectangle (51, 8);
        
        \filldraw[fill=black!20,draw=black] (111,0) rectangle (111+1, 2);
        \filldraw[fill=black!20,draw=black] (111,6) rectangle (111+1, 8);
        
        \node[rotate=270,anchor=south] at (111, 4) {Grating};

        \draw[line width=2,dotted, red] (0,0) -- (361, 0) node[midway] (sample) {};
        
        \draw[<-, thick] (sample) -- ($(sample)+(0,3)$) -- ($(sample)+(30,3)$) node[anchor=west] (){Sample};
        %\draw[line width=2,dotted, red] (0,8) -- (361, 8);
        
        %\draw[line width=2] (70, 0) -- (70, 32) node[rotate=90, anchor=north, midway] {Source};
        
        \filldraw[black] (71,4) circle (2pt) node[rotate=270, anchor=south] {Source};
        
         %\useasboundingbox (bound.south east) rectangle (bound.north west); %So that axes don't push the image right
        
        \tikzset{x=2,y=2};
        \draw [->,thick] (-4, -4) -- (-4, 5) node[anchor=east] {\(H_y\)};
        \draw [->, thick] (-4, -4) node{\(\odot\)} node[anchor=north east]{\(E_z\)} -- (5, -4) node[anchor=north west,shift={(-3,0)}] {\(H_x\)};
    \end{tikzpicture} 
    \label{fig:grating_pbc}
    }
    \end{minipage}
    \par\bigskip
    \def\dy{0.01cm} 
    \begin{minipage}{\columnwidth}
        \subfloat[]{
        \begin{tikzpicture}[x = \dx, y = \dy]
            \draw[fill=blue!20, opacity=0.5] (0,0) rectangle (361, 537) ; % PML rectangle
            \draw[fill=white] (40,40) rectangle (361-40, 537-40) ; % Fill in white area
            \node[] at (361/2, 537-20) {PML};
            \node[] at (361/2, 20) {PML};
            \node[rotate=270] at (20, 537/2) {PML};
            \node[rotate=270] at (361-20, 537/2) {PML};
        
            \node[draw,anchor=south west,  minimum width=361*\dx, minimum height=537*\dy] (bound) at (0,0) {};
            
            \filldraw[fill=black!20,draw=black] (111,0) rectangle (111+1, 2);
            
            %% Grating
            \foreach \y in {-10,...,25}
                \filldraw[fill=black!20,draw=black] (111,408-2 - 8*\y) rectangle (111+1, 408+2 - 8*\y);
            
            %%
            
            %\filldraw[fill=blue!20,draw=blue,opacity=0.5] (0,0) rectangle (40, 457) node[pos=.5, rotate=270, opacity=1] {PML};
            %\filldraw[fill=blue!20,draw=blue,opacity=0.5] (361,0) rectangle (361-40, 457) node[pos=.5, rotate=270, opacity=1] {PML};
            %\filldraw[fill=blue!20,draw=blue,opacity=0.5] (0,0) rectangle (361, 40) node[pos=.5, opacity=1] {PML};
            %\filldraw[fill=blue!20,draw=blue,opacity=0.5] (0,457-40) rectangle (361, 457) node[pos=.5, opacity=1] {PML};

            \draw[line width=2,dotted, red] (0,408) -- (361, 408) node[pos=0.65, below, black] {\textbf{TF/SF Boundary}};
            
            \draw[decoration={brace,mirror,raise=5pt},decorate,line width=1pt] (111, 408-2 + 8*10) -- (111, 408-8*25) node[midway,xshift=-14pt,rotate=270] {Grating};
            
            \draw[|-|,xshift=6pt, shorten <= 2pt] (111, 408-2) -- (111, 408-8*25) node[midway,anchor=south,rotate=270] {\(N^E+0.5\)};
            \draw[|-|,xshift=6pt, shorten >= 2pt] (111, 408 + 8*10) -- (111, 408+2) node[midway,anchor=south,rotate=270] {10};

            %\draw[<-,very thick, shorten <= 2pt] (250, 408) |- (280, 560) node[anchor=west] {SF Region};
            
            \node[] at (270, 408-60-40) {TF Region};
            \node[] at (270, 408+60) {SF Region};
            
            %\draw[->, very thick,green] (180, 300) -- (180, 200);
            
            %%
            \tikzset{x=2,y=2};
            \draw [->,thick] (-4, -4) -- (-4, 5) node[anchor=east] {\(H_y\)};
            \draw [->, thick] (-4, -4) node{\(\odot\)} node[anchor=north east]{\(E_z\)} -- (5, -4) node[anchor=north west,shift={(-3,0)}] {\(H_x\)};
        \end{tikzpicture}
        \label{fig:grating_edge}
        }
    \end{minipage}
    %\vspace{10pt}
    \caption{Schematic of (a) a single unit cell simulation and (b) of the edge effects simulation of the subwavelength grating. The red dotted line in (a) displays where electric and magnetic fields are sampled. The red dotted line in (b) represents the TF/SF boundary at which the recorded fields are injected into the simulation. The ``total-field'' (TF) and ``scattered-field'' (SF) regions are denoted. Ten periods of the grating appear in the SF region, while \(N^E+0.5\) periods of the grating are in the TF region.}
\end{figure}

First, a simulation of a single unit cell of the grating was carried out, with PBCs (see \cref{fig:grating_pbc}). The operating frequency of the structure was set to \(f_\text{op} = \SI{12}{\giga\hertz}\). The simulation was run for \(2^{12}\) steps.

The dimensions were set to \(d_x=\SI{4.37}{\cm}\) and \(d_y=\SI{1.25}{\mm}\) in the \(x\) and \(y\) directions respectively. The perfect electric conductor (PEC) making up the metal of the grating spanned half of the unit cell and was one Yee cell thick.

The domain was discretized into 361 and 889 Yee cells in the \(x\) and \(y\) directions respectively (\(\Delta x=\Delta y=\lambda/160\)) in order to finely model the subwavelength features of the fields and grating. A Courant number of \(0.9\) was chosen to calculate the time step. 

The boundaries parallel to the \(x\) axis were lined by 40-cell-thick perfectly-matched layers (PMLs) \cite{taflove2005computational}. The boundaries parallel to the \(y\) axis were bordered by PBCs. 

The source excitation was defined as a Gaussian electric current \(J_z\) containing the operating frequency. The source was positioned \(\lambda/4\) units to the side of the grating. The ASM was used to record fields along one boundary parallel to \(x\) (denoted by a red dashed line on the simulated unit cell in \cref{fig:grating_pbc}) as a function of time.

Symmetry was exploited to improve ASM efficiency (see \cref{sec:asm}). Fields from this simulation were used to approximate the internal region of the grating, away from the edges.

Another simulation was then carried out to approximate the fields by the edges (see \cref{fig:grating_edge}). The discretization of space and time of this simulation matches that of the PBC simulation described above, as does the dimension parallel to the \(x\) axis, the positioning of the grating and the PML parameters. 

The simulation of the edge had PMLs along each boundary. Ten unit cells of the grating were placed between the TF/SF boundary and the PML, in the SF region. The number of periodic unit cells in the SF region does not appear to greatly change the field patterns in the TF (total-field) region. A variable number of unit cells of the grating were located in the TF region. Empty space spanning \SI{5}{\cm} was included in the \(y\) direction, shown below the grating in \cref{fig:grating_edge}.

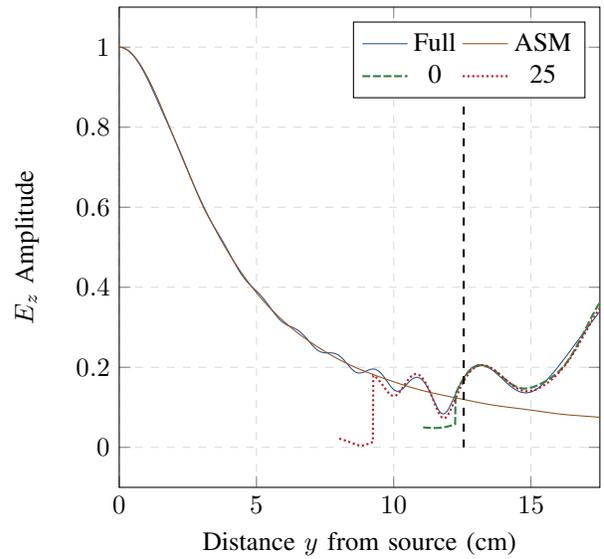
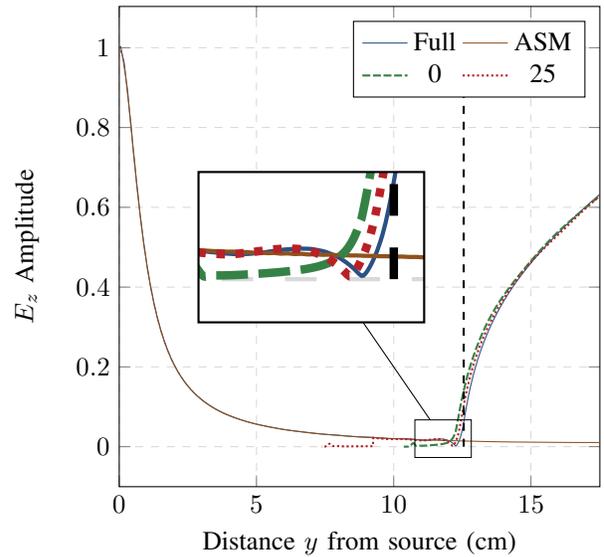
\begin{figure}
    \centering
    \begin{minipage}{\columnwidth}
    \subfloat[]{
    \label{fig:2d_sample_l}
    \begin{tikzpicture}
          \begin{axis}[
              width=0.9\columnwidth, % Scale the plot to \linewidth
              height=0.9\columnwidth,
              grid=major, 
              grid style={dashed,gray!30},
              xmin=0,
              xmax=17.51,
              %ymin=-35,
              %xlabel=X Axis $U$, % Set the labels
              ylabel={\(E_z\) Amplitude},
              xlabel={Distance \(y\) from source (cm)},
              x filter/.code={\pgfmathparse{#1*100}\pgfmathresult},
    %          x unit=\si{\volt}, % Set the respective units
    %          y unit=\si{\ampere},
              %legend style={at={(0.5,-0.2)},anchor=north},
              x tick label style={rotate=0,anchor=north},
              legend pos=north east,
              ylabel style={alias=ylab},
              yticklabel style={text width=2em,align=right},
              %cycle list name=color list,
              cycle list name=color_pattern,
              legend columns=2, 
            ]
            \addplot+[] table[x=x,y=e,col sep=comma] {2d/lambda/full.csv};
            \addplot+[] table[x=x,y=e,col sep=comma] {2d/lambda/asm.csv};
            \addplot+[restrict x to domain=10.9902461755208:20] table[x=x,y=e,col sep=comma] {2d/lambda/0.csv};
            \addplot+[restrict x to domain=7.99232159552083:20] table[x=x,y=e,col sep=comma] {2d/lambda/25.csv};

            \addplot +[mark=none, dashed, color=black] coordinates {(+0.1255, 0) (+0.1255, 1)};
            
            \legend{Full, ASM, 0, 25}
          \end{axis}
    \end{tikzpicture}
    }
    \end{minipage}
    \par\bigskip
    \begin{minipage}{\columnwidth}
    \subfloat[]{
    \label{fig:2d_sample_l/20}
    \begin{tikzpicture}[spy using outlines={rectangle, magnification=4,connect spies}]
          \begin{axis}[
              width=0.9\columnwidth, % Scale the plot to \linewidth
              height=0.9\columnwidth,
              grid=major, 
              grid style={dashed,gray!30},
              x filter/.code={\pgfmathparse{#1*100}\pgfmathresult},
              xmin=0,
              xmax=17.51,
              %xmax=0.1751,
              %ymin=-35,
              %xlabel=X Axis $U$, % Set the labels
              ylabel={\(E_z\) Amplitude},
              xlabel={Distance \(y\) from source (cm)},
    %          x unit=\si{\volt}, % Set the respective units
    %          y unit=\si{\ampere},
              %legend style={at={(0.5,-0.2)},anchor=north},
              x tick label style={rotate=0,anchor=north},
              legend pos=north east,
              ylabel style={alias=ylab},
              yticklabel style={text width=2em,align=right},
              %cycle list name=color list,
              cycle list name=color_pattern,
              legend columns=2, 
            ]
            \addplot+[] table[x=x,y=e,col sep=comma] {2d/lambda20/full.csv};
            \addplot+[] table[x=x,y=e,col sep=comma] {2d/lambda20/asm.csv}; 
            \addplot+[] table[x=x,y=e,col sep=comma] {2d/lambda20/0.csv};
            \addplot+[] table[x=x,y=e,col sep=comma] {2d/lambda20/25.csv}; 
            %\addplot+[ ] table[x=x,y=e,col sep=comma] {2d/lambda20/test.csv}; 

            %\addplot+[] table[x=x,y=e,col sep=comma] {2d/lambda20/del.csv}; 
            \addplot +[mark=none, dashed, color=black] coordinates {(+0.1255, 0) (+0.1255, 1)};
            
            \legend{Full, ASM, 0, 25}
            
            \coordinate (spypoint) at (axis cs:11.8,0.02);
            \coordinate (spyviewer) at (axis cs:7,0.5);
            \spy[width=3cm,height=2cm] on (spypoint) in node [fill=white] at (spyviewer);
            
          \end{axis}
    \end{tikzpicture}
    }
    \end{minipage}
    \caption{Normalized magnitudes of frequency-domain \(E_z\) field samples at (a) \(\lambda\) units from the grating and (b) \(\lambda/20\) units from the grating. Each plot shows fields from full-sized simulations of a 200-unit-cell grating (labelled in the legend as ``Full''), fields obtained from the ASM, and fields captured in the simulation of edge unit cells. In the latter case, the legend entry is the number of edge unit cells simulated (\(N^E\)). A vertical dashed line marks the end of the grating. An inset shows a magnification of the region near the edge of the grating. The transitions of the fields of the edge-cell simulations from the SF region, where fields are low, to the TF regions, where fields track those of the full simulation, are distinguishable.}
    \label{fig:2d_samples}
\end{figure}

Sampling was performed along lines positioned at \(\lambda\) and \(\lambda/20\) units away from the grating (on the opposite side of the source) and Fourier transformed at the operating frequency. Fields obtained from a finite simulation, from the ASM simulation (i.e. a simulation of an infinitely-periodic grating) and from the grating edge simulation are shown in \cref{fig:2d_samples}. The simulations of the edges were carried out with 25 edge unit cells and without any edge unit cells. Fields to either side of the TF/SF boundary are included in these plots.

Edge effects and the beginnings of side lobes are visible in both plots. There are strong ripples around the edges \cref{fig:2d_sample_l} in particular. Edge fields can be accurately estimated with very few, or even no edge unit cells. However, if one wants to smoothly merge the fields obtained by the ASM and the fields obtained via the edge unit cells simulation to produce a single continuous field, several unit cells may need to be included in the edge unit cells simulation as the ASM may be insufficient to describe fields near the edges. In \cref{fig:2d_sample_l}, for example, the ASM fields may be merged more readily with the 25-unit-cell simulation of the edge than with the 0-unit-cell simulation.

%When many edge cell unit cells are simulated, there remain slight differences between the ASM fields and those found from the simulations of the edges. This is apparently due to the absence of interaction between the two edges on either side of the grating. 

\subsection{Computational Cost Analysis}\label{sec:2d-timing}

The times taken to complete the full-sized simulations, the ASM simulations and the edge unit cells simulations were recorded. Due to symmetry, only half of the grating was simulated. A perfect magnetic conductor (PMC) was placed along the symmetry axis to produce the desired field symmetry profile.

\Cref{tab:2d_timing} describes the timing of the ASM and edge unit cell simulations when \(N^E=0\) and \(N^E=25\) edge unit cells are considered. The time taken to simulate the entire structure is included as well. The full simulation time assumes ASM parallelization.

\begin{table*}
    \centering
    \caption{Time taken to simulate a 200-unit-cell subwavelength grating using the proposed algorithm (varying edge unit cells \(N^E\)) and a full simulation of the finite structure. The ASM integration order used for estimating the internal fields away from the edges is provided along with the runtime of each simulation. }
    \label{tab:2d_timing}
    \begin{tabular}{
    l
    S[table-text-alignment = center]
    S[table-text-alignment = center]
    S[table-text-alignment = center]
    S[table-text-alignment = center]
    S
    }\toprule
        {Sim. Type} & {\(N^E\)} & {ASM Integration Order} & {ASM Sim. Time (s)} & {Edge Sim. Time (s)} & {Total Sim. Time (s)} \\\midrule
        \multirow{2}{4em}{Edge} & 0 & 213 & 1.52 & 26.4 & 27.9\\
         & 25 & 201 & 1.52 & 46.6 & 48.1\\\midrule
        Full & {---} & {---} & {---} & {---} & 129\\
         \bottomrule
    \end{tabular}
\end{table*}

\section{3D Finite-Sized Structure Simulations}

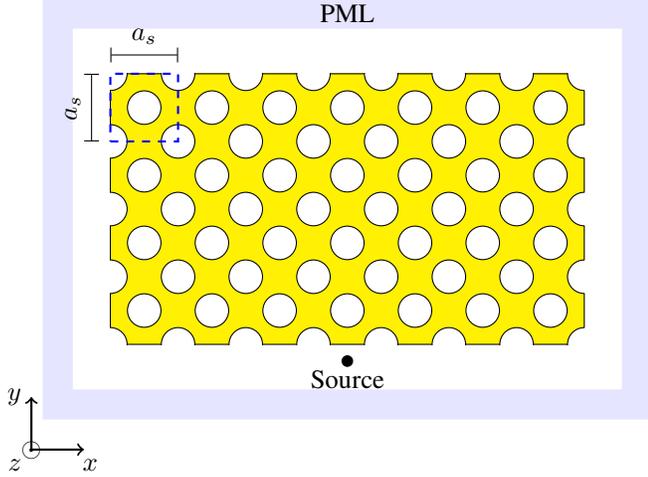
\begin{figure}
    \centering
    \def \dx {0.05cm}
    \def \dy {0.05cm}
    \def \ncell {7}
    \begin{tikzpicture}[x = \dx, y = \dy,
    invclip/.style={clip,insert path={{[reset cm]
        (-16383.99999pt,-16383.99999pt) rectangle (16383.99999pt,16383.99999pt)}}} ]
        
    \draw[blue!20, opacity=0.5, line width=4mm]     (-\ncell*2*\dx, -16*\dy) rectangle (\ncell*20*\dx, 88*\dy) ;
    \node[] at (\ncell*18*\dx/2, 88*\dy) {PML};
    
    \begin{scope}
        \begin{pgfinterruptboundingbox}
            \foreach \x in {0,...,9}
            \foreach \y in {0,...,4}
            {
                \clip[invclip] (18*\x,18*\y) circle (4.4548);
                \clip[invclip] ({18*(\x+0.5)},{18*(\y+0.5)}) circle (4.4548);
            }
        \end{pgfinterruptboundingbox}
        
        \draw[fill=yellow] (0,0) rectangle (\ncell*18*\dx, 72*\dy);
    \end{scope}
    
    \begin{scope}
    \begin{pgfinterruptboundingbox}
        \clip (0,0) rectangle (\ncell*18*\dx, 72*\dy);
    \end{pgfinterruptboundingbox}
    
    \foreach \x in {0,...,9}
            \foreach \y in {0,...,4}
            {
                \draw[fill=white] (18*\x,18*\y) circle (4.4548);
                \draw[fill=white] ({18*(\x+0.5)},{18*(\y+0.5)}) circle (4.4548);
            }
    \end{scope}

    \draw[dashed, gray,thick, blue] (0, 72*\dy) rectangle (18*\dx, 72*\dy-18*\dx); %around a period
    
    \draw[|-|,transform canvas={yshift=+2.5mm}] (0, 72*\dy) -- (18*\dx, 72*\dy) node[midway,yshift=+7pt,rotate=0] {\(a_s\)};
    \draw[|-|,transform canvas={xshift=-2.5mm}] (0, 72*\dy) -- (0, 72*\dy-18*\dy) node[midway,xshift=-7pt,rotate=90] {\(a_s\)};
    
    %% Source
    
    \draw[fill=black] (\ncell*18*\dx/2, -4.45*\dx) circle (2pt) node[anchor=north] {Source};

    %%
    %\tikzset{x=2,y=2};
    \draw [->,thick] (-\ncell*3, -\ncell*4) -- (-\ncell*3, -\ncell*2) node[anchor=east] {\(y\)};
    \draw [->, thick] (-\ncell*3, -\ncell*4) node{\(\odot\)} node[anchor=north east]{\(z\)} -- (-\ncell*1, -\ncell*4) node[anchor=north west,shift={(-3,0)}] {\(x\)};
    \end{tikzpicture}
    \caption{A schematic of the full-sized simulation of the three-dimensional photonic crystal from \cite{luo2002all} (\(x\)-\(y\) plane, \(z=0\)). }
    \label{fig:3d_photonic}
\end{figure}

Consider a structure which is periodic in the \(x\) and \(z\) directions in a three-dimensional simulation, as in \cref{fig:3d_photonic}. The algorithm outlined below is used to estimate the fields throughout the structure following the two-step process described previously (\cref{sec:methodology}).

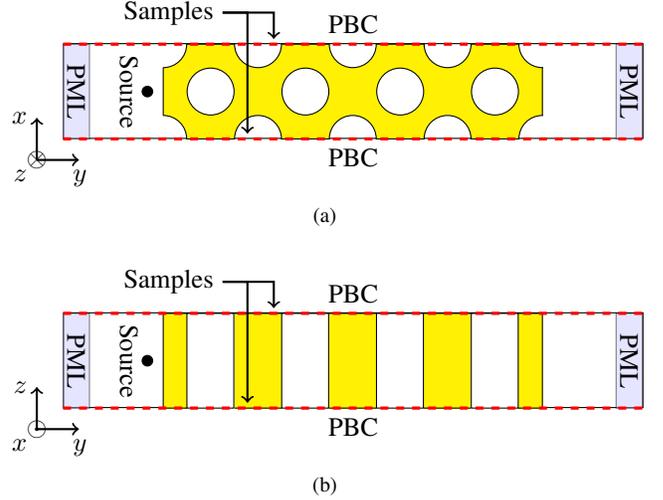
\begin{figure}
    \centering
    \def \dx {0.07cm}
    \def \dy {0.07cm}
    \begin{minipage}{\columnwidth}
        \subfloat[]{
            \begin{tikzpicture}[x = \dx, y = \dy, invclip/.style={clip,insert path={{[reset cm]
                (-16383.99999pt,-16383.99999pt) rectangle (16383.99999pt,16383.99999pt)}}} ]
                    
                \node[draw,anchor=south west, minimum width=110*\dx, minimum height=18*\dy, label=below:PBC, label=above:PBC] (bound) at (-55*\dx, -9*\dy) {};
                
                \begin{scope}
                \begin{pgfinterruptboundingbox}
                    \foreach \x in {0,...,4} {
                        \clip[invclip] (-4*9 + 18*\x+9,0) circle (4.4548);
                        \clip[invclip] (-4*9 + 18*\x,9) circle (4.4548);
                        \clip[invclip] (-4*9 + 18*\x,-9) circle (4.4548);
                    }
                \end{pgfinterruptboundingbox}
                \draw[fill=yellow] (-4*9*\dx, -9*\dy) rectangle (4*9*\dx, 9*\dy);
                \end{scope}
                
                \begin{scope}
                \begin{pgfinterruptboundingbox}
                    \clip (-4*9*\dx, -9*\dy) rectangle (4*9*\dx, 9*\dy);
                
                \end{pgfinterruptboundingbox}
                \foreach \x in {0,...,4} {
                        \draw (-4*9 + 18*\x+9,0) circle (4.4548);
                        \draw (-4*9 + 18*\x,9) circle (4.4548);
                        \draw (-4*9 + 18*\x,-9) circle (4.4548);
                        }
                \end{scope}
                
                \draw[fill=blue!20,opacity=0.5] (-50*\dx, -9*\dy) rectangle (-55*\dx, 9*\dy) node[midway, rotate=270,opacity=1] {PML};
                \draw[fill=blue!20,opacity=0.5] (50*\dx, -9*\dy) rectangle (55*\dx, 9*\dy) node[midway, rotate=270,opacity=1] {PML};
                
                \draw[very thick, dashed, red] (-55*\dx,9*\dy) -- (55*\dx,9*\dy);
                \draw[very thick, dashed, red] (-55*\dx,-9*\dy) -- (55*\dx,-9*\dy);
                
                \node[] at (-35*\dx, 15*\dy) (sample) {Samples};
                \draw[->, shorten >= 2pt, thick] (sample) -- (-20*\dx,15*\dy) -- (-20*\dx,-9*\dy) {};
                \draw[->, shorten >= 2pt, thick] (sample) -- (-15*\dx, 15*\dy) -- (-15*\dx, 9*\dy);

                \draw[fill=black] (-39, 0) circle (2pt) node[anchor=north,rotate=270] {Source};
        
                \draw [->,thick] (-60, -13) -- (-60, -5) node[anchor=east] {\(x\)};
                \draw [->, thick] (-60, -13) node{\(\otimes\)} node[anchor=north east]{\(z\)} -- (-52, -13) node[anchor=north west,shift={(-3,0)}] {\(y\)};
            
            \end{tikzpicture}
            \label{fig:3d_ze0}
    }
    \end{minipage}
    \par\bigskip
    \begin{minipage}{\columnwidth}
        \subfloat[]{
                \begin{tikzpicture}[x = \dx, y = \dy]
                    
                \node[draw,anchor=south west, minimum width=110*\dx, minimum height=18*\dy, label=below:PBC, label=above:PBC] (bound) at (-55*\dx, -9*\dy) {};
                
                \draw[fill=yellow] (-4*9*\dx, -9*\dy) rectangle (4*9*\dx, 9*\dy);
                
                \foreach \x in {0,...,3} {
                        \draw[fill=white] (-4*9*\dx + 18*\x*\dx+4.5*\dx, -9*\dy) rectangle ++(4.4548*2, 18*\dy);
                    }
                
                %%%%%
                \draw[fill=blue!20,opacity=0.5] (-50*\dx, -9*\dy) rectangle (-55*\dx, 9*\dy) node[midway, rotate=270,opacity=1] {PML};
                \draw[fill=blue!20,opacity=0.5] (50*\dx, -9*\dy) rectangle (55*\dx, 9*\dy) node[midway, rotate=270,opacity=1] {PML};
                
                \draw[very thick, dashed, red] (-55*\dx,9*\dy) -- (55*\dx,9*\dy);
                \draw[very thick, dashed, red] (-55*\dx,-9*\dy) -- (55*\dx,-9*\dy);
                
                \node[] at (-35*\dx, 15*\dy) (sample) {Samples};
                \draw[->, shorten >= 2pt, thick] (sample) -- (-20*\dx,15*\dy) -- (-20*\dx,-9*\dy) {};
                \draw[->, shorten >= 2pt, thick] (sample) -- (-15*\dx, 15*\dy) -- (-15*\dx, 9*\dy);

                \draw[fill=black] (-39, 0) circle (2pt) node[anchor=north,rotate=270] {Source};
        
                \draw [->,thick] (-60, -13) -- (-60, -5) node[anchor=east] {\(z\)};
                \draw [->, thick] (-60, -13) node{\(\odot\)} node[anchor=north east]{\(x\)} -- (-52, -13) node[anchor=north west,shift={(-3,0)}] {\(y\)};
            
            \end{tikzpicture}
        \label{fig:3d_xe0}
    }
    \end{minipage}
    \caption{(a) The \(z=0\) plane and (b) the \(x=0\) plane of one unit cell of the three-dimensional photonic crystal. The ASM is used to estimate the fields in the inner region of the finite structure and to record the fields which are used to estimate the fields along the edges of the structure. The faces along which fields are sampled in the simulated unit cell are marked by dashed red lines.}
    \label{fig:3d_1_period}
\end{figure}

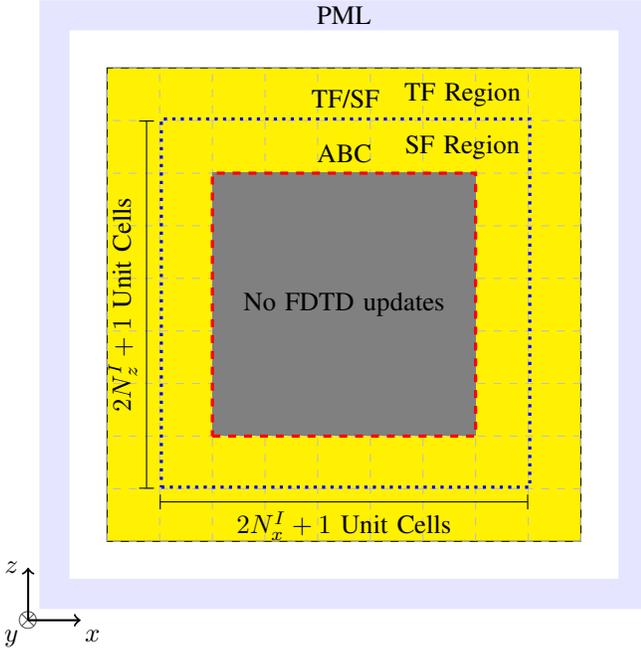
\begin{figure}
    \centering
    \def \dx {0.025cm}
    \def \dy {0.025cm}
    \def \ncell {14}
    \begin{tikzpicture}[x = \dx, y = \dy]
    
    \draw[blue!20, opacity=0.5, line width=4mm] (-\ncell*11, -\ncell*11) rectangle (\ncell*11, \ncell*11) ;
    \node[] at (0, \ncell*11) {PML};
    %\node[draw, blue!20, opacity=0.5, line width=4mm,anchor=south west, minimum width=22*\ncell*\dx, minimum height=22*\ncell*\dy] at (-\ncell*11, -\ncell*11) {} ;
    
    \draw[fill=yellow] (-\ncell*9, -\ncell*9) rectangle (\ncell*9, \ncell*9);
    
    %Don't go by 2s to make it denser. But there should be 1 period between ABC and TF/SF
    \foreach \x in {0,2,...,18}
    {
        \draw[lightgray, dashed] (-\ncell*9 + \x*\ncell, -\ncell*9) -- (-\ncell*9 + \x*\ncell, \ncell*9);
        \draw[lightgray, dashed] (\ncell*9, -\ncell*9+\x*\ncell) -- (-\ncell*9, -\ncell*9+\x*\ncell);
    }

     \node[draw, rectangle, very thick, dotted, blue, anchor=south west, minimum width=14*\ncell*\dx, minimum height=14*\ncell*\dy, label=above:TF/SF] at (-\ncell*7, -\ncell*7) {};

    \draw[|-|] (-\ncell*7, -\ncell*7.5) -- (\ncell*7, -\ncell*7.5) node[midway, below] {\(2N^I_x+1\) Unit Cells};
    \draw[|-|] (-\ncell*7.5, -\ncell*7) -- (-\ncell*7.5, \ncell*7) node[midway, above, rotate=90] {\(2N^I_z+1\) Unit Cells};
    
     \node[fill=gray,anchor=south west, minimum width=10*\ncell*\dx, minimum height=10*\ncell*\dy, label=above:ABC] at (-\ncell*5, -\ncell*5) {} node[] {No FDTD updates};
     
     \node[] at (\ncell*4.5, \ncell*6) {SF Region};
     \node[] at (\ncell*4.5, \ncell*8) {TF Region};
     
     %\draw[fill=black] (0,0) circle (2pt) node[anchor=north] {Source};
     
     %\node[draw, rectangle, red, dashed, very thick, anchor=south west, minimum width=10*\ncell*\dx, minimum height=10*\ncell*\dy, label=above:ABC] at (-\ncell*5, -\ncell*5) {};
     
     \draw[rectangle, red, dashed, very thick] (-\ncell*5, -\ncell*5) rectangle (\ncell*5, \ncell*5);

    %%
    %\tikzset{x=2,y=2};
    \draw [->,thick] (-\ncell*12, -\ncell*12) -- (-\ncell*12, -\ncell*10) node[anchor=east] {\(z\)};
    \draw [->, thick] (-\ncell*12, -\ncell*12) node{\(\otimes\)} node[anchor=north east]{\(y\)} -- (-\ncell*10, -\ncell*12) node[anchor=north west,shift={(-3,0)}] {\(x\)};
    
    \end{tikzpicture}
    \caption{Schematic of the simulation of the edges of the three-dimensional photonic crystal (\(x\)-\(z\) plane, \(y = -2a_s-r\)). The photonic crystal is shown in yellow, and each unit cell is outlined in grey dotted lines. The simulation domain comprises of (from inside to outside): (1) a block (grey) which is not updated by FDTD, (2) an ABC (red dashed line) encasing the non-updated block, (3) a TF/SF boundary (blue dotted line) several unit cells from the ABC, (4) the photonic crystal boundary and (5) the PML along the simulation boundary (blue).  The ``total-field'' (TF) and ``scattered-field'' (SF) regions are denoted.}
    \label{fig:3d_edges}
\end{figure}

First, a single unit cell is simulated with PBCs along the four periodic faces, and absorbers along the other two (aperiodic) faces (see \cref{fig:3d_1_period}). Let us define the inner region of the structure, where edge effects are negligible, to be \(N^I_x\) unit cells to either side of the center unit cell in the \(x\) direction, and \(N^I_z\) unit cells on either side of the center unit cell in the \(z\) direction (see \cref{fig:3d_edges}; the center unit cell, with the source, appears in \cref{fig:3d_photonic}). The ASM can then be used to record tangential fields along the boundary of the inner region. That is, the ASM can be used to record:
\begin{alignat}{2}
    &U_x(x,y,(N^I_z+1/2)d_z,t), &&U_y(x,y,(N^I_z+1/2)d_z,t)\\
    &U_x(x,y,-(N^I_z+1/2)d_z,t),\quad&&U_y(x,y,-(N^I_z+1/2)d_z,t)
\end{alignat}
where \(|x| \le (N^I_x+1/2)d_x\) and 
\begin{alignat}{2}
   &U_y((N^I_x+1/2)d_x,y,z,t),  &&U_z((N^I_x+1/2)d_x,y,z,t)\\
    &U_y(-(N^I_x+1/2)d_x,y,z,t), \quad &&U_z(-(N^I_x+1/2)d_x,y,z,t)
\end{alignat}
where \(|z| \le (N^I_z+1/2)d_z\). In these expressions, \(U_i\) denotes both electric and magnetic field components.

The recorded fields are then used to estimate the fields near the edges. As shown in \cref{fig:3d_edges}, the recorded fields are injected several unit cells away from the edges in the periodic directions. An absorber should be placed behind the TF/SF boundary in the SF region to prevent reflections of fields in the SF region back into the TF region. As with the 2D simulation (\cref{sec:subwavelength_grating}), some space should be left between the TF/SF boundary and the absorber as a buffer zone for the absorber and to facilitate accuracy quantification of the algorithm (see \cref{sec:quantifying_convergence}). The absorber in the SF region should not be a standard PML, since convex PMLs are not stable \cite{teixeira1999causality}.

In the region enclosed by the absorber, fields do not need to be simulated since the ASM simulation estimates the fields within this region. No updates are required within this region. %Whenever the region which is not updated is large compared to the rest of the simulation, there become significant computational savings in runtime and memory requirements. When this region is relatively small, the simulation cost approaches that of a standard, full-sized simulation of the structure.

%Number of unit cells behind and after the TFSF boundary vs. time

%What to do about the fact that there are many edges?

\subsection{Numerical Results: Finite-Sized 3D Photonic Crystal}\label{sec:photonic_crystal}

%0.5 GHz -> 600 THz (1.2e6)
%59.96e-2 m -> 499.7 nm (8.334e-7)

A 3D model of a periodic photonic crystal was simulated to illustrate the procedure of edge-effects determination. The photonic crystal simulated is notable due to its exhibition of negative refraction at all angles of incidence, making it useful in lensing applications \cite{luo2002all}.

The photonic crystal was composed of a dielectric slab with a relative permittivity \(\varepsilon_r = 12.0\) centered at the origin, with staggered rows of vacuum-cylinders (radius \(r=\SI{33.6}{\nano\meter}\)) in the \(z\)-direction (see \cref{fig:3d_photonic}). The period in the \(x\) and \(y\) directions was \(a_s=\SI{135.7}{\nano\meter}\). Since the infinite structure was uniform in the \(z\) direction in the ASM simulation (see \cref{fig:3d_xe0}), the period was chosen to be \(a_s\). The photonic crystal was four periods thick in the \(y\) direction, while the size of the crystal may be varied in the other axes. Throughout this section, the photonic crystal studied had \(11\) unit cells in both the \(x\) and \(z\) directions.

The slab was excited by a Gaussian magnetic current point source \(M_z\) located at a distance \(r\) away from the slab face (i.e. at \(y = -2a_s-r\)) in the \(x=z=0\) plane. The excitation spanned frequencies between \(f_\text{min} = \SI{0}{\peta\hertz}\) and \(f_\text{max} = \SI{1.2}{\peta\hertz}\) in order to enclose the operating frequency \(f_\text{op} = \SI{600}{\tera\hertz}\) (cyan visible light).

The domain was discretized as \(\Delta x = \Delta y =  \SI{8.58}{\nm}\) in the \(x\) and \(y\) directions. The \(z\) axis, along which the photonic crystal is uniform, was discretized in intervals of \(\Delta z = \SI{22.8}{\nm}\). 

A single unit cell was simulated using the ASM to calculate the fields in the internal region (see \cref{fig:3d_1_period}). In the simulation of the edge effects, the inner region (in which no FDTD updates occur) was surrounded by first-order Mur's absorbing boundary conditions (ABC) \cite{taflove2005computational}. The border of the ABC block was located one period \(a_s\) from the TF/SF sources in the \(x\) and \(z\) directions in all simulations, as illustrated in \cref{fig:3d_edges}. The number of periodic unit cells in the SF region between the ABC and the TF/SF boundary does not appear to greatly change the field patterns in the TF region. 

The simulation region was terminated with PMLs (thickness \SI{12.5}{\nano\meter}), located \SI{25.0}{\nano\meter} from the edges of the photonic crystal.

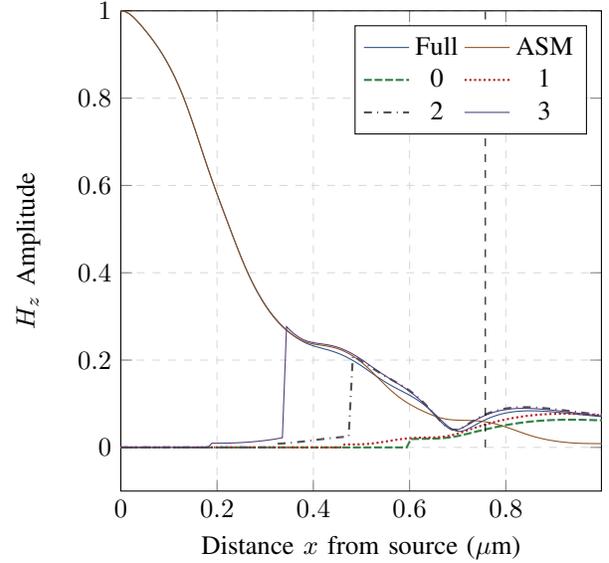
\begin{figure}
    \begin{tikzpicture}%[spy using outlines={rectangle, magnification=2,connect spies}]
          \begin{axis}[
              width=0.9\columnwidth, % Scale the plot to \linewidth
              height=0.9\columnwidth,
              grid=major, 
              grid style={dashed,gray!30},
              x filter/.code={\pgfmathparse{#1*0.833}\pgfmathresult},
              xmin=0,
              xmax=0.998581039749008,
              %xmax=0.1751,
              %ymin=-1,
              %xlabel=X Axis $U$, % Set the labels
              ymax=1,
              ylabel={\(H_z\) Amplitude},
              xlabel={Distance \(x\) from source (\(\mu\)m)},
    %          x unit=\si{\volt}, % Set the respective units
    %          y unit=\si{\ampere},
              %legend style={at={(0.5,-0.2)},anchor=north},
              x tick label style={rotate=0,anchor=north},
              legend pos=north east,
              ylabel style={alias=ylab},
              yticklabel style={text width=2em,align=right},
              %cycle list name=color list,
              cycle list name=color_pattern,
              legend columns=2, 
            ]
            
            \addplot+[solid] table[x=x,y=e,col sep=comma] {3d/11/full.csv};
            \addlegendentry{Full}
            
            \addplot+[solid] table[x=x,y=e,col sep=comma] {3d/asm.csv};
            \addlegendentry{ASM}

           % \addplot+[solid,restrict x to domain=0.7575:1] table[x=x,y=e,col sep=comma] {3d/11/0.csv};
            \addplot+[] table[x=x,y=e,col sep=comma] {3d/11/0.csv};
            \addlegendentry{0}
            
            %\addplot+[solid,restrict x to domain=0.6198:1] table[x=x,y=e,col sep=comma] {3d/11/1.csv};
            \addplot+[] table[x=x,y=e,col sep=comma] {3d/11/1.csv};
            \addlegendentry{1}
            
            %\addplot+[solid,restrict x to domain= 0.48207:1] table[x=x,y=e,col sep=comma] {3d/11/2.csv};
            \addplot+[] table[x=x,y=e,col sep=comma] {3d/11/2.csv};
            \addlegendentry{2}
            
            %\addplot+[solid] table[x=x,y=e,col sep=comma,restrict x to domain=0.34:1] {3d/11/3.csv};
            \addplot+[] table[x=x,y=e,col sep=comma] {3d/11/3.csv};
            \addlegendentry{3}
            
            %\pgfplotsset{cycle list shift=-4}
                        
            %\addplot+[densely dashed] table[x=x,y=e,col sep=comma] {3d/11/0.csv};
            %\addplot+[densely dashed] table[x=x,y=e,col sep=comma] {3d/11/1.csv};
            %\addplot+[densely dashed] table[x=x,y=e,col sep=comma] {3d/11/2.csv};
            %\addplot+[densely dashed] table[x=x,y=e,col sep=comma,restrict x to domain=0:0.35] {3d/11/3.csv};
            
            \draw[dashed] (axis cs:0.7575, 0) -- (axis cs:0.7575, 1);

            %\legend{Full, ASM, 0, 1, 2, 3}

            %\coordinate (spypoint) at (axis cs:0.8,0.1);
            %\coordinate (spyviewer) at (axis cs:0.7,-0.5);
            %\spy[width=0.7\columnwidth,height=1.5cm] on (spypoint) in node [fill=white] at (spyviewer);
            
            %\draw[<-,very thick] (axis cs:0.27, 0.04) |- (axis cs:0.4, 0.35);
          \end{axis}
    \end{tikzpicture}
    \caption{Normalized magnitudes of Fourier-transformed \(H_z\) samples at a distance \(r\) from the photonic crystal. The plot shows fields from full-sized simulations of the finite structures (labelled in the legend as ``Full''), fields obtained from the ASM, and fields captured in the simulation of edge unit cells. In the latter case, the legend entry is the number of edge unit cells simulated (\(N^E_x=N^E_z\)). The transitions of the fields of the edge-cell simulations from the SF region, where fields are low, to the TF regions, where fields are higher, are clearly distinguishable in the \(N^I=2\) and \(N^I=3\) cases.}
    \label{fig:3d_samples}
\end{figure}

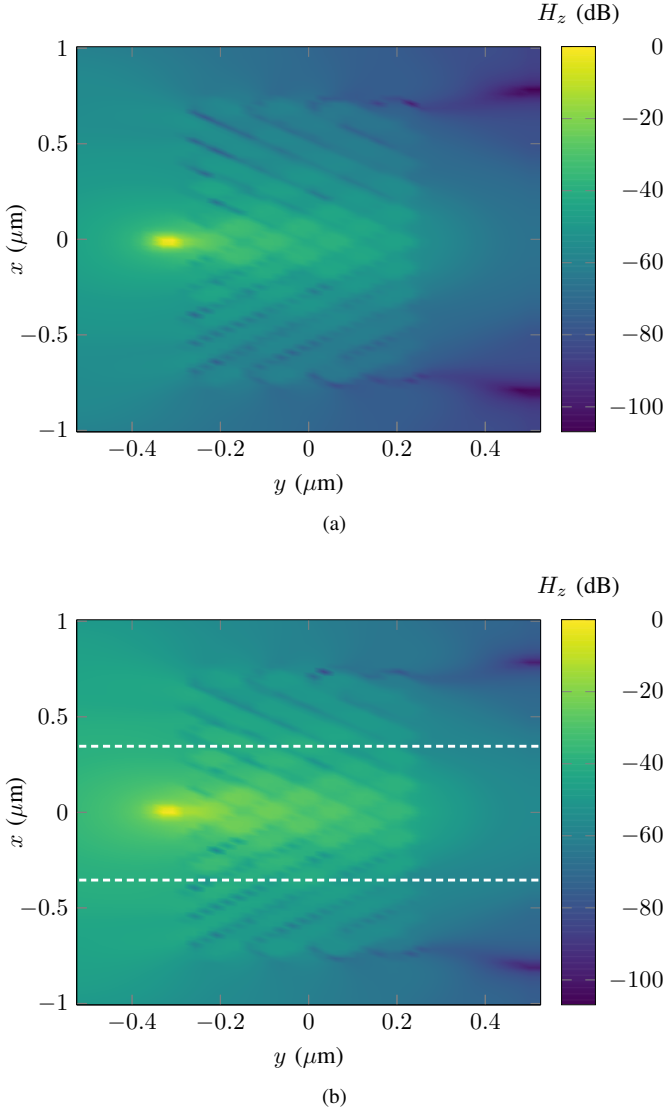
\begin{figure}
    \centering
    \begin{minipage}{\columnwidth}
    \subfloat[]{
    \label{fig:3d_slice_full}
    \centering
    \begin{tikzpicture}[scale = 0.9]
    \begin{axis}[view={0}{90},colormap/viridis,
                ylabel={\(x\) (\(\mu\)m)},
                xlabel={\(y\) (\(\mu\)m)},
                ylabel shift={-10pt},
                colorbar,
                colorbar style={
                    title={\(H_z\) (dB)},
                    yticklabel style={
                        text width=2.5em,
                        align=right,
                        }
                    }]
            \addplot3[surf,  mesh/rows=83,   shader = interp] table[x=x, y=y, z=z, col sep = comma] {3d/surf/full.csv};
        \end{axis}
        
        %%
        %%\tikzset{x=5,y=5};
        %\draw [->,thick] (-0.75, -0.75) -- (-0.75, 0.0) node[anchor=east] {\(y\)};
        %\draw [->, thick] (-0.75, -0.75) node{\(\odot\)} node[anchor=north east]{\(z\)} -- (0.0, -0.75) node[anchor=north west] {\(x\)};
    \end{tikzpicture}
    }
    \end{minipage}
    \par\bigskip
    \begin{minipage}{\columnwidth}
    \subfloat[]{
    \label{fig:3d_slice_merged}
    \begin{tikzpicture}[scale = 0.9]
        \begin{axis}[view={0}{90},colormap/viridis,
                ylabel={\(x\) (\(\mu\)m)},
                xlabel={\(y\) (\(\mu\)m)},
                ylabel shift={-10pt},
                colorbar,
                colorbar style={
                    title={\(H_z\) (dB)},
                    yticklabel style={
                        text width=2.5em,
                        align=right,
                        }
                    },
                    point meta min=-106.8979,
                    ]
                \addplot3[surf,  mesh/rows=83,   shader = interp] table[x=x, y=y, z=z, col sep = comma] {3d/surf/merged.csv};
                \draw[densely dashed, very thick, white] (axis cs:-0.65455, -0.3545) -- (axis cs:0.65455,-0.3545);
                \draw[densely dashed, very thick, white] (axis cs:-0.65455,0.3458) -- (axis cs:0.65455,0.3458);
        \end{axis}
        
        %%
        %%\tikzset{x=5,y=5};
        %\draw [->,thick] (-0.75, -0.75) -- (-0.75, 0.0) node[anchor=east] {\(y\)};
        %\draw [->, thick] (-0.75, -0.75) node{\(\odot\)} node[anchor=north east]{\(z\)} -- (0.0, -0.75) node[anchor=north west] {\(x\)};
    \end{tikzpicture}
    }
    \end{minipage}
    \caption{The Fourier transform of the normalized \(H_z\) field amplitudes at \(f_\text{op}\) on the \(z=0\) plane determined from (a) a full-sized simulation and (b) from merging fields from the ASM simulation with fields from the simulation of the edges (\(N^E_x=N^E_z=3\) edge unit cells). White dashed lines in (b) outline the boundary between the internal region and edge regions.}
    \label{fig:3d_slice}
\end{figure}

The Courant number was set to \(0.9\) and the simulation was run for \(2^{9}\) time steps. A sample of \(H_z\) at \((x,2a_s+\SI{36.5}{\nano\meter},0)\)---lying along the image plane of the source \cite{luo2002all}---was measured and Fourier transformed at the operating frequency \(f_\text{op}\). The sample magnitudes at various numbers of edge unit cells is shown in \cref{fig:3d_samples}, and are compared against the full-sized simulation of the finite photonic crystal. Fields to either side of the TF/SF boundary are shown in these plots. \(H_z\) fields from full finite simulations and from the algorithm described here are presented throughout the entire \(z=0\) plane in \cref{fig:3d_slice}.

Good correspondence exists between the fields from the full-sized simulations versus fields combined from the ASM simulations and from the simulation of the edge unit cells, when two or more edge unit cells are simulated. In particular, the simulations of the edges are able to approximate fields outside the slab. Simulations with PBCs alone cannot do that meaningfully. 

\subsection{Computational Cost Analysis}
The time taken to complete the full-sized simulations, the ASM simulations and the edge unit cells simulations of an \(11\times 11\) photonic crystal are recorded in \cref{tab:3d_timing}. The ASM integration order in the periodic directions (\(x\) and \(z\)) and the time taken to simulate each unit cell are noted in the table. The ASM integration orders are not the same in each dimension, due to discretization discrepancies. As expected, significant time savings are possible when the ASM simulations are parallelized. 

\begin{table*}
    \centering
    \caption{Time taken to simulate an \(11\times 11\) photonic crystal using the proposed algorithm (varying edge unit cells \(N^E\)) and a full simulation of the finite structure. The ASM integration order used for estimating the internal fields away from the edges is provided along with the runtime of each simulation. }
    \label{tab:3d_timing}
    \begin{tabular}{
    l
    S[table-text-alignment = center]
    S[table-text-alignment = center]
    S[table-text-alignment = center]
    S[table-text-alignment = center]
    S
    }\toprule
        {Sim. Type} & {\(N^E\)} & {ASM Integration Order (\(x\times z\))} & {ASM Sim. Time (s)} & {Edge Sim. Time (s)} & {Total Sim. Time (s)} \\\midrule
        \multirow{4}{4em}{Edge} & 0 & {\(26\times23\)}  & 28.0  & 340  & 368 \\
          & 1 & {\(25\times22\)}  & 27.8 & 458  & 486 \\
          & 2 & {\(24\times21\)}  & 27.6  &  627 & 655 \\
          & 3 & {\(23\times20\)}  & 27.3  &  816 & 843 \\
         \midrule
        Full & {---} & {---} & {---} & {---} & 1400\\
         \bottomrule
    \end{tabular}
\end{table*}

\section{Quantifying Convergence of Fields Generated by the Proposed Algorithm versus Fields from Full Finite Simulations}
\label{sec:quantifying_convergence}

As seen in the previous sections, the merger of the ASM fields and fields from simulations of the edge unit cells can approximate fields from the full simulations of the finite structures. The accuracy of the approximation may vary as the number of unit edge unit cells \(N^E\) changes. A metric for numerically quantifying this accuracy is therefore desired.

A criterion for quantifying convergence may be identified from the previous results. A smooth transition between the fields of the ASM solution and the fields from the edge simulation is required for accurately merging fields. %The field amplitude in the SF region in the edge unit cell simulation should be small. Fields in the SF region arise since the fields produced by the TF/SF boundary are those of the infinite case, but the edge simulation is finite.

\Cref{fig:2d_sample_l} may be used to illustrate this criterion. There is no smooth transition between the ASM solution and the \(N^E=0\) case. When \(N^E=25\), there is a smoother transition between the edge simulation fields and the ASM fields.

To quantify the convergence criterion, the Euclidean (\(L_2\)) norm is used to describe the difference between the fields determined by the ASM simulations and the fields deduced from the edge simulations over several unit cells.

Consider a 2D simulation. Define \(E_\text{ASM}(y) = E_z^\text{ASM}(\lambda, y, f_\text{op})\) and \(E_{N^E}(y) = E_z^{N^E}(\lambda, y, f_\text{op})\) where \(E_z^\text{ASM}(x,y,f)\) and \(E_z^{N^E}(x,y,f)\) are the \(z\)-directed electric fields (in frequency-domain) determined from the ASM simulation and from simulations of \(N^E\) edges, respectively. Then, the difference between \(E_\text{ASM}\) and \(E_{N^E}\) over \(P_\text{TF}\) unit cells in the total-field (TF) region past the TF/SF boundary (located at \(y = (0.5+N^I)d_y\)) may be expressed using the formula:
\begin{align}
\begin{split}
    ||E_\text{ASM} &- E_{N^E}||_2^\text{TF} =\\ &\sqrt{\int_{(0.5+N^I)d_y}^{(P_\text{TF}+0.5+N^I)d_y} |E_\text{ASM} - E_{N^E}|^2 \,dy}.
\end{split}
\end{align}
The 3D formula may be derived in the same fashion.

%The variable \(P_\text{SF}\) should be set to the number of unit cells between the TF/SF boundary and the absorber in the SF region. 
The variable \(P_\text{TF}\) should be chosen to span the length over which agreement between the edge-cell simulation and ASM is desired. If \(P_\text{TF}\) is too small, \(||E_\text{ASM} - E_{N^E}||_2^\text{TF}\) may suggest convergence in certain cases even if it has not yet been reached. If \(E_\text{ASM}\) and \(E_{N^E}\) happen to be similar around a small domain past the TF/SF boundary in the TF region, the metric will suggest convergence, even if it has not been reached.

The convergence metric was applied to the fields of the the subwavelength grating sampled \(\lambda\) units away from the grating (\cref{sec:subwavelength_grating}) and of the fields of the photonic crystal (\cref{sec:photonic_crystal}) described above. The results are presented in \cref{tab:convergence_grating,tab:convergence_pc}. In the results, cases with the largest number of edge unit cells have the lowest scores, implying strongest convergence. As anticipated, \(||E_\text{ASM} - E_{N^E}||_2^\text{TF}\) becomes more conservative as \(P_\text{TF}\) is increased. More precisely, the metric \(||E_\text{ASM} - E_{N^E}||_2^\text{TF}\) (as a function of \(N^E\)) tends to zero more slowly when when \(P_\text{TF}\) is increased.

\begin{table}[]
    \centering
    \caption{Convergence quantification of a 200-unit cell subwavelength grating (see \cref{sec:subwavelength_grating}) as the number edge unit cells \(N^E\) varies. The data in each column is normalized to the largest value. }
    \label{tab:convergence_grating}
    %\sisetup{table-number-alignment = right}
    \begin{tabular}{
    S%[table-text-alignment = center]
    S%[table-text-alignment = center]
    S%[table-text-alignment = center]
    S%[table-text-alignment = center]
    }\toprule
        {\(N^E\)} & 
        {\thead{\(||E_\text{ASM} - E_{N^E}||_2^\text{TF}\) \\ \(P_\text{TF}=1\)}} & 
        {\thead{\(||E_\text{ASM} - E_{N^E}||_2^\text{TF}\) \\ \(P_\text{TF}=10\)}}  \\
        \midrule
        0 & 1 & 1  \\
        25 & 0.169 & 0.294  \\
         \bottomrule
    \end{tabular}
\end{table}
% 0 raw values:
%7.4287e-04
%0.0085
%0.0057

\begin{table}[]
    \centering
    \caption{Convergence quantification of an \(11\times11\) photonic crystal (see \cref{sec:subwavelength_grating}) as the number edge unit cells \(N^E\) varies. The data in each column is normalized to the largest value.}
    \label{tab:convergence_pc}
    %\sisetup{table-number-alignment = right}
    \begin{tabular}{
    S%[table-text-alignment = center]
    S%[table-text-alignment = center]
    S%[table-text-alignment = center]
    S%[table-text-alignment = center]
    }\toprule
        {\(N^E\)} & 
        {\thead{\(||E_\text{ASM} - E_{N^E}||_2^\text{TF}\) \\ \(P_\text{TF}=1\)}} & 
        {\thead{\(||E_\text{ASM} - E_{N^E}||_2^\text{TF}\) \\ \(P_\text{TF}=2\)}}\\
        \midrule
        0 & 0.608 & 1  \\
        1 & 1 & 0.932 \\
        2 & 0.386 & 0.372 \\
        3 & 0.0297 & 0.261\\
         \bottomrule
    \end{tabular}
\end{table}
% 0 cells
% 0.0101
% 0.0234 <-
% 0.0105 <-

% 1 cells
% 0.0166 <-
% 0.0218
% 0.0057

% 2 cells
% 0.0064
% 0.0087
% 0.0072

% 3 cells
% 4.9299e-04
% 0.0061
% 0.0053

\section{Convergence of Fields on Finite Periodic Structures}
Finite periodic structures approximate infinitely periodic structures increasingly well as the number of unit cells increases. Typically, the number of unit cells needed for convergence is determined by monitoring simulations of increasingly large finite structures \cite{holter2002size, zhou2008size, parsons2011copper}. The present algorithm may be used to determine the number of unit cells required for convergence more efficiently. By selecting the inner region of unit cells \(2N^I+1\) to be the region where convergence is desired, the number of edge unit cells \(N^E\) can be increased from zero until convergence is reached.

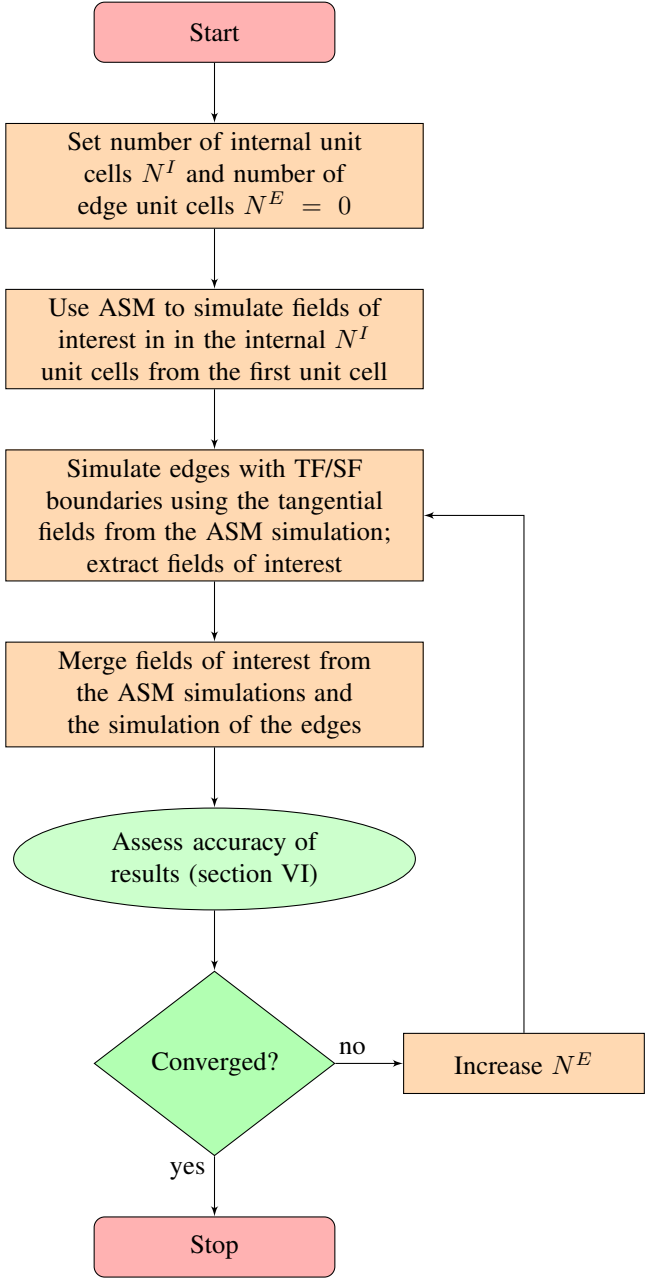
\begin{figure}
    \centering
    %%%%%    
    \begin{tikzpicture}[
    line/.style = {draw, -latex'},
    node distance = 8mm and 9mm,
      start chain = A going below,
      base/.style = {draw, minimum width=32mm, minimum height=8mm,
                     align=center, on chain=A},
 startstop/.style = {base, rectangle, rounded corners, fill=red!30},
         cloud/.style = {base, draw, ellipse,fill=green!20,},
   process/.style = {base, rectangle, fill=orange!30},
        io/.style = {base, trapezium, 
                     trapezium left angle=70, trapezium right angle=110,
                     fill=blue!30},
  decision/.style = {base, diamond, fill=green!30},
  every edge quotes/.style = {auto=right}]
        % Place nodes
        \node [startstop] (start) {Start};
        \node [process,text width=0.6\columnwidth] (seta) {Set number of internal unit cells \(N^I\) and number of edge unit cells \(N^E=0\)};
        \node [process,text width=0.6\columnwidth] (asm)        {Use ASM to simulate fields of interest in in the internal \(N^I\) unit cells from the first unit cell
        };
        \node [process,text width=0.6\columnwidth] (edges)         {Simulate edges with TF/SF boundaries using the tangential fields from the ASM simulation; extract fields of interest };
        \node [process,text width=0.6\columnwidth]  (merge)       {Merge fields of interest from the ASM simulations and the simulation of the edges};
        \node [cloud,text width=0.4\columnwidth] (accuracy)       {Assess accuracy of results (\cref{sec:quantifying_convergence})};
        \node [decision] (converged)       {Converged?};
        \node [startstop] (stop) {Stop};
        \node [process, right=of converged] (gpp) {Increase \(N^E\)};
        
    %%%
    \draw[line] (start) -- (seta);
    \draw[line] (seta) -- (asm);
    \draw[line] (asm) -- (edges);
    \draw[line] (edges) -- (merge);
    \draw[line] (merge) -- (accuracy);
    \draw[line] (accuracy) -- (converged);
    \draw[line] (converged) -- (stop) node [near start,anchor=east] {yes};
    \draw[line] (converged) -- (gpp) node [near start,anchor=south] {no};
    \draw[line] (gpp) |- (edges);
    \end{tikzpicture}
    
    \caption{A flowchart describing the procedure of efficiently determining how large a finite periodic structure must be for fields to behave as though the structure were infinite over a region of interest.}
    \label{fig:convergence_flowchart}
\end{figure}

Suppose convergence is desired within a range of \(2N^I+1\) unit cells in the periodic direction (as \cref{fig:2d_edges}). In other words, it is desired that the fields within the region of \(2N^I+1\) internal unit cells agree with the fields generated by the infinitely-periodic structure (at a given distance from the grating).

The ASM---which assumes infinite periodicity---may be used to simulate the fields on the \(2N^I+1\) internal unit cells, since convergence has been assumed in this region. Then, as described above, TF/SF boundaries can be used to inject these fields onto a simulation of a finite number of unit cells. This simulation, which is much smaller than a full-sized simulation of the finite structure, may be run several times with increasingly more unit cells until convergence is reached. Note that this algorithm requires one set of ASM simulations. This iterative procedure is outlined schematically in \cref{fig:convergence_flowchart}.

\subsection{Numerical Results: Subwavelength Grating}

The algorithm above was applied to analyze the convergence of the fields of a finite subwavelength grating to its infinite periodic counterpart. The characteristics and simulation details of the grating are described in \cref{sec:subwavelength_grating}. 
The ASM was used to simulate \(N^I=29\) periodic unit cells (i.e. \(1.45\lambda_\text{op}\)) to either side of the center unit cell. The fields generated by the ASM were used to simulate edge unit cells as described in \cref{sec:2d_edge_effects}. \(E_z\) samples were taken one wavelength \(\lambda\) away from the grating and were Fourier transformed at the operating frequency.

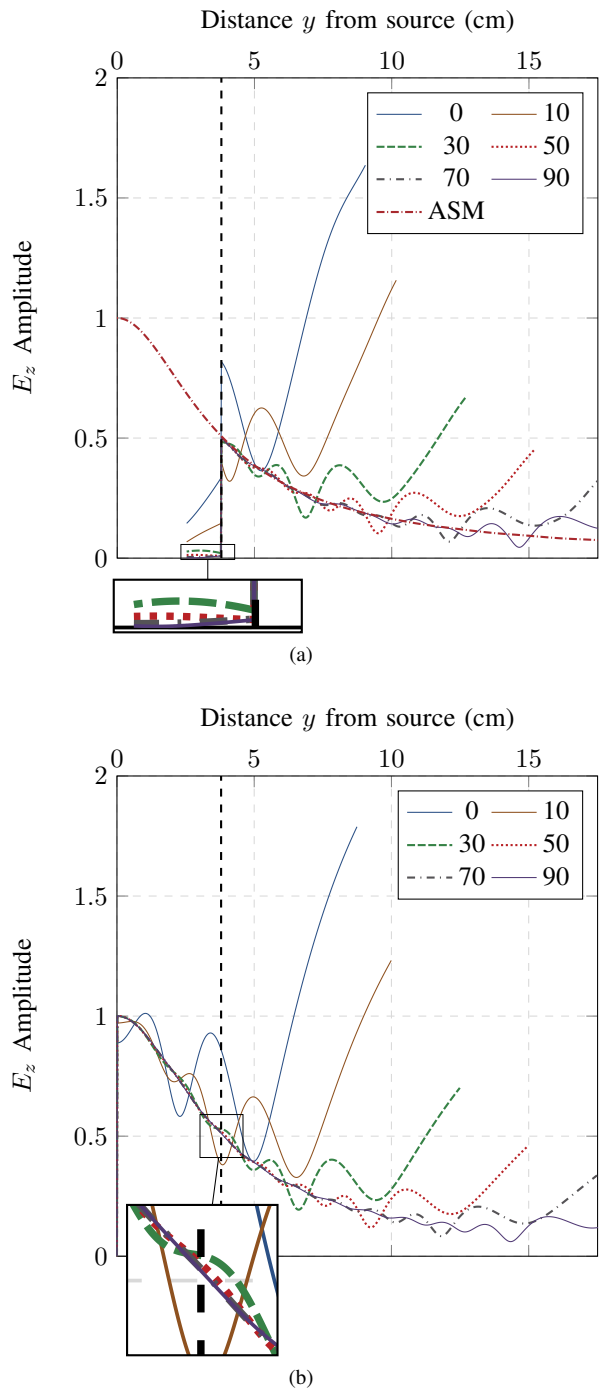
\begin{figure}
    \centering
    \begin{minipage}{\columnwidth}
    \subfloat[]{
    \label{fig:2d_convergence_edges}
    \begin{tikzpicture}[spy using outlines={rectangle, magnification=3.5,connect spies}]
          \begin{axis}[
              width=0.9\columnwidth, % Scale the plot to \linewidth
              height=0.9\columnwidth,
              grid=major, 
              grid style={dashed,gray!30},
              xmin=0,
              xmax=17.51,
              ymin=0,
              ymax=2,
              %xlabel shift={-10pt},
              %xlabel=X Axis $U$, % Set the labels
              ylabel={\(E_z\) Amplitude},
              xlabel={Distance \(y\) from source (cm)},
              x filter/.code={\pgfmathparse{#1*100}\pgfmathresult},
    %          x unit=\si{\volt}, % Set the respective units
    %          y unit=\si{\ampere},
              %legend style={at={(0.5,-0.2)},anchor=north},
              x tick label style={rotate=0,anchor=south},
              legend pos=north east,
              yticklabel style={text width=2em,align=right},
              %cycle list name=color list,
              cycle list name=color_pattern,
              legend columns=2, 
              xlabel near ticks,
              xticklabel pos=top,
            ]
            \addplot+[restrict x to domain=2.5298:9.0562] table[x=x,y=e,col sep=comma] {converge/0.csv};
            \addplot+[restrict x to domain=2.5298:10.18] table[x=x,y=e,col sep=comma] {converge/10.csv};
            \addplot+[restrict x to domain=2.5298:12.679] table[x=x,y=e,col sep=comma] {converge/30.csv};
            \addplot+[restrict x to domain=2.5298:15.177] table[x=x,y=e,col sep=comma] {converge/50.csv};
            \addplot+[restrict x to domain=2.5298:17.675] table[x=x,y=e,col sep=comma] {converge/70.csv};
            \addplot+[restrict x to domain=2.5298:20.174] table[x=x,y=e,col sep=comma] {converge/90.csv};
            
            \addplot+[] table[x=x,y=e,col sep=comma] {converge/asm.csv};
            
            \addplot +[mark=none, dashed, color=black] coordinates {(0.037942, 0) (0.0379420, 2)};
            \legend{0,10,30,50,70,90,ASM}
            
            \coordinate (spypoint) at (axis cs:3.3,0.025);
            %\coordinate (spyviewer) at (axis cs:2,1.5);
            \coordinate (spyviewer) at (axis cs:3.3,-0.2);
            \spy[width=2.5cm,height=0.70cm] on (spypoint) in node [fill=white] at (spyviewer);
          \end{axis}
          
          %\useasboundingbox (bound.south east) rectangle (bound.north west);
    \end{tikzpicture}
    }
    \end{minipage}
    \par\bigskip
    \begin{minipage}{\columnwidth}
    \subfloat[]{
    \label{fig:2d_convergence_full}
    \begin{tikzpicture}[spy using outlines={rectangle, magnification=3.5,connect spies}]
          \begin{axis}[
              width=0.9\columnwidth, % Scale the plot to \linewidth
              height=0.9\columnwidth,
              grid=major, 
              grid style={dashed,gray!30},
              xmin=0,
              xmax=17.51,
              ymin=0,
              ymax=2,
              %xlabel shift={-10pt},
              %xlabel=X Axis $U$, % Set the labels
              ylabel={\(E_z\) Amplitude},
              xlabel={Distance \(y\) from source (cm)},
              x filter/.code={\pgfmathparse{#1*100}\pgfmathresult},
    %          x unit=\si{\volt}, % Set the respective units
    %          y unit=\si{\ampere},
              %legend style={at={(0.5,-0.2)},anchor=north},
              x tick label style={rotate=0,anchor=south},
              legend pos=north east,
              yticklabel style={text width=2em,align=right},
              %cycle list name=color list,
              cycle list name=color_pattern,
              legend columns=2, 
              xlabel near ticks,
              xticklabel pos=top,
            ]
            \addplot+[] table[x=x,y=e,col sep=comma] {converge/full/0.csv};
            \addplot+[] table[x=x,y=e,col sep=comma] {converge/full/10.csv};
            \addplot+[] table[x=x,y=e,col sep=comma] {converge/full/30.csv};
            \addplot+[] table[x=x,y=e,col sep=comma] {converge/full/50.csv};
            \addplot+[] table[x=x,y=e,col sep=comma] {converge/full/70.csv};
            \addplot+[] table[x=x,y=e,col sep=comma] {converge/full/90.csv};
            %Norm factor: 1/4.542e-15 
            %\legend{0 (75.159160), 10 (88.408533), 30 (117.308786), 50 (151.960240), 70 (181.142928), 90 (213.570256)}
            
            \addplot +[mark=none, dashed, color=black] coordinates {(0.037942, 0) (0.037942, 2)};

            \legend{0, 10, 30, 50, 70, 90}
            
            \coordinate (spypoint) at (axis cs:3.8,0.5);
            %\coordinate (spyviewer) at (axis cs:2,1.5);
            \coordinate (spyviewer) at (axis cs:3.1,-0.1);
            \spy[width=2cm,height=2cm] on (spypoint) in node [fill=white] at (spyviewer);
          \end{axis}
    \end{tikzpicture}
    }
    \end{minipage}
    
    \caption{Normalized magnitudes of frequency-domain \(E_z\) field samples at \(\lambda\) units from the grating (\(N^I=29\)) as the number of edge unit cells \(N^E\) is varied. Figure (a) illustrates how convergence (\(\pm N^I\) unit cells from the center unit cell) may be obtained by selecting \(N^E\) high enough such that the edge simulation fields agree with the ASM fields. An inset (bottom) magnifies weak fields in the SF region. Full simulations of finite gratings of various sizes are shown in (b). Numbers in the legends refer to the number of edge unit cells \(N^E\). An inset (bottom) magnifies fields at the boundary between the inner unit cells and the boundary unit cells. The vertical dashed lines in (a) and (b) denote the boundary between the inner unit cells and the boundary unit cells. }
    \label{fig:2d_convergence}
\end{figure}

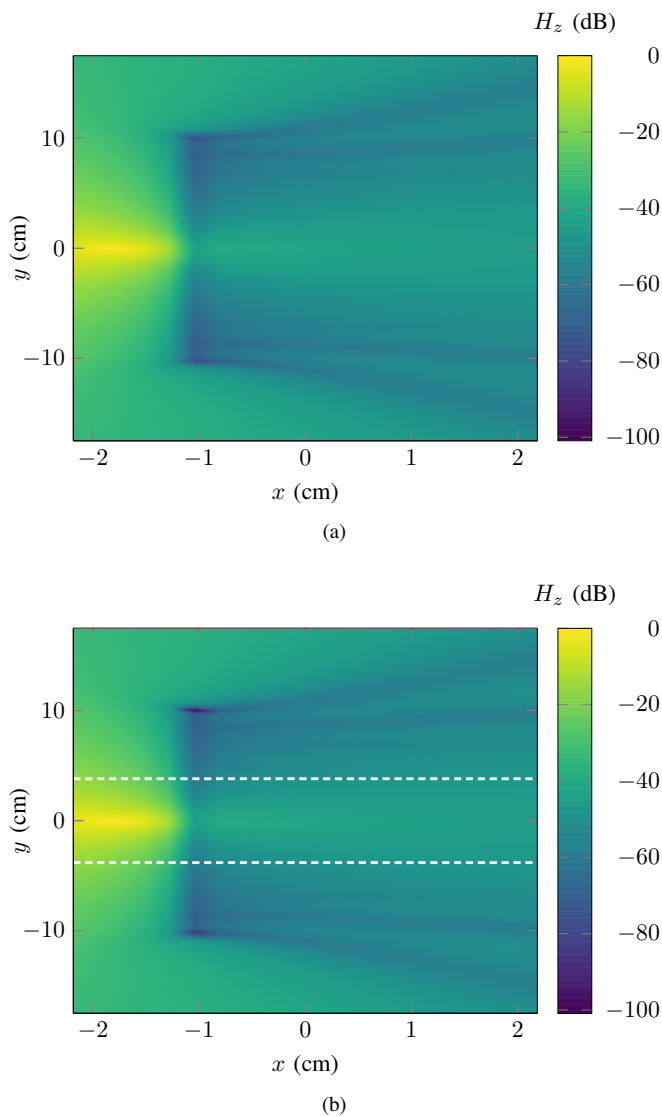
\begin{figure}
    \centering
    \begin{minipage}{\columnwidth}
    \subfloat[]{
    \label{fig:2d_slice_full}
    \centering
    \begin{tikzpicture}[scale = 0.9]
    \begin{axis}[view={0}{90},colormap/viridis,
                ylabel={\(y\) (cm)},
                xlabel={\(x\) (cm)},
                ylabel shift={-10pt},
                colorbar,
                colorbar style={
                    title={\(H_z\) (dB)},
                    yticklabel style={
                        text width=2.5em,
                        align=right,
                        }
                    },
                    point meta min=-100.9258,
                    x filter/.code={\pgfmathparse{#1*100}\pgfmathresult},
                    y filter/.code={\pgfmathparse{#1*100}\pgfmathresult},]
            \addplot3[surf,  mesh/rows=113,   shader = interp] table[x=x, y=y, z=z, col sep = comma] {converge/surf/full.csv};
        \end{axis}
        
        %%
        %%\tikzset{x=5,y=5};
        %\draw [->,thick] (-0.75, -0.75) -- (-0.75, 0.0) node[anchor=east] {\(y\)};
        %\draw [->, thick] (-0.75, -0.75) node{\(\odot\)} node[anchor=north east]{\(z\)} -- (0.0, -0.75) node[anchor=north west] {\(x\)};
    \end{tikzpicture}
    }
    \end{minipage}
    \par\bigskip
    \begin{minipage}{\columnwidth}
    \subfloat[]{
    \label{fig:2d_slice_merged}
    \begin{tikzpicture}[scale = 0.9]
        \begin{axis}[view={0}{90},colormap/viridis,
                ylabel={\(y\) (cm)},
                xlabel={\(x\) (cm)},
                ylabel shift={-10pt},
                colorbar,
                colorbar style={
                    title={\(H_z\) (dB)},
                    yticklabel style={
                        text width=2.5em,
                        align=right,
                        }
                    },
                     x filter/.code={\pgfmathparse{#1*100}\pgfmathresult},
                     y filter/.code={\pgfmathparse{#1*100}\pgfmathresult},]
                \addplot3[surf,  mesh/rows=114,   shader = interp] table[x=x, y=y, z=z, col sep = comma] {converge/surf/merged.csv};
                \draw[densely dashed, very thick, white] (axis cs:-2.1850, -3.8099) -- (axis cs:2.1850,-3.8099);
                \draw[densely dashed, very thick, white] (axis cs:-2.1850, 3.8099) -- (axis cs:2.1850,3.8099);
        \end{axis}
        
        %%
        %%\tikzset{x=5,y=5};
        %\draw [->,thick] (-0.75, -0.75) -- (-0.75, 0.0) node[anchor=east] {\(y\)};
        %\draw [->, thick] (-0.75, -0.75) node{\(\odot\)} node[anchor=north east]{\(z\)} -- (0.0, -0.75) node[anchor=north west] {\(x\)};
    \end{tikzpicture}
    }
    \end{minipage}
    \caption{The Fourier transform of the normalized \(E_z\) field amplitudes at \(f_\text{op}\) determined from (a) a full-sized simulation and (b) from merging fields from the ASM simulation (\(N^I=29\) internal unit cells) with fields from the simulation of the edges (\(N^E=30\) edge unit cells). White dashed lines in (b) outline the boundary between the internal region and edge regions. Observe how the edges are distant enough from the internal region that the edge effects do not enter the internal region.}
    \label{fig:2d_slice}
\end{figure}

\Cref{fig:2d_convergence_edges} shows the field samples from the ASM simulation and from the simulation of the edge unit cells. Values of the convergence metrics (\cref{sec:quantifying_convergence}) are tabulated in \cref{tab:convergence} for various values of \(N^E\). Based on these results, reasonable convergence appears at around \(N^E=30\), corresponding to a grating of \(119\) unit cells in total. The plots in \cref{fig:2d_convergence_edges} visually suggest as well that convergence is attained at around 30 edge unit cells. Convergence at this point was confirmed by a full simulation of finite-sized grating shown in \cref{fig:2d_convergence_full}. The \(N^E=30\) case is presented in \cref{fig:2d_slice}, which compares the fields over the entire computational domain from the full simulation of the finite grating with the fields determined from the algorithm given here. 

\begin{table}[]
    \centering
    \caption{Convergence quantification of a subwavelength grating \(N^I=29\) as the number of edge unit cells \(N^E\) varies. The data in each column is normalized to the largest value.}
    \label{tab:convergence}
    %\sisetup{table-number-alignment = right}
    \begin{tabular}{
    S%[table-text-alignment = center]
    S%[table-text-alignment = center]
    S%[table-text-alignment = center]
    }\toprule
        {\(N^E\)} & {\thead{\(||E_\text{ASM} - E_{N^E}||_2^\text{TF}\) \\ \(P_\text{TF} = 10\)}}  \\ \midrule
        0 & 1  \\
        10 & 0.480  \\
        30 & 0.0402  \\
        50 & 0.0292  \\
        70 & 0.0325 \\
        90 & 0.0294  \\
         \bottomrule
    \end{tabular}
\end{table}

Note that since sampling was performed one wavelength away from the grating (orthogonal to the direction of periodicity), convergence of fields within the region of interest was assured at distance of at most one wavelength from the grating.

\subsection{Computational Cost Analysis}

The time taken to simulate the edges in each configuration is provided in \cref{tab:2d_converge_timing}, along with the corresponding time required to simulate the entire grating. The table also includes the time taken to run full-sized simulations at the various sizes. As in \cref{sec:2d-timing}, symmetry was exploited both in the ASM simulations and in the full-size simulations. As expected, there are significant runtime savings at high ASM parallelization rates.

\begin{table*}
    \centering
    \caption{A comparison of the time taken to simulate the edge unit cells of a subwavelength grating (\(N^I=29\) internal unit cells) versus the time taken to simulate the corresponding finite structure.}
    \label{tab:2d_converge_timing}
    \begin{tabular}{
    S[table-text-alignment = center]
    S[table-text-alignment = center]
    S[table-text-alignment = center]
    S[table-text-alignment = center]
    S[table-text-alignment = center]
    S
    }\toprule
        {\(N^E\)} & {ASM Integration Order} & {ASM Sim. Time (s)} & {Edge Sim. Time (s)} & {Total Sim. Time (s)} & {Full Structure Sim. Time (s)} \\\cmidrule(lr){1-1}
        \cmidrule(lr){2-5}
        \cmidrule(lr){6-6}
        0 & {\multirow{6}{4em}{178}}  & {\multirow{6}{4em}{1.52}} & 29.0  & 30.5 & 51.0 \\
        10 & & & 36.1 & 37.6 & 60.0 \\
        30 & & & 52.0 & 53.5 & 80.9 \\
        50 & & & 75.7 & 79.2 & 102 \\
        70 & & & 89.3 & 90.8 & 125 \\
        90 & & & 110 & 112 & 148 \\
        \bottomrule
    \end{tabular}
\end{table*}

\section{Summary and Conclusions}
This paper presents a new FDTD technique to simulate finite periodic structures. The technique involves a two-step procedure: First, a simulation of a single unit cell terminated by PBCs is performed. The ASM method is used to remove unwanted image sources generated by the PBCs. Electric and magnetic fields tangential to the periodic faces are then sampled several unit cells away from the source as functions of time. Second, a simulation of the edges of the structure is carried out. A TF/SF boundary is used to inject the fields recorded from the ASM simulation onto the edge unit cells. When the number of edge unit cells in the simulation is large enough, there is a smooth transition between fields from the ASM simulation and fields from the simulation of the edges. A metric for quantifying the success of the algorithm is suggested.

Surfaces in two-dimensional simulations have two disjoint ends, while three-dimensional surfaces have borders along four contiguous edges. They thus are handled in different ways.

This paper's algorithm was used to determine the fields of a subwavelength grating in two-dimensions and to determine the fields of a photonic crystal in three-dimensions. When the ASM simulation was parallelized, the algorithm gave rise to computational savings.

The algorithm described here may also be used to efficiently determine the number of unit cells required to attain convergence within a certain region of the periodic structure. The ASM can be used to estimate the fields within the convergent inner region. Then, successive simulations of more and more edge unit cells may be used to analyze convergence.

Further optimizations of the algorithm described here are possible. For instance, ASM integration orders used in the examples above are conservative and may be relaxed in many cases.

The structures considered here have identical unit unit cells within their inner region and edge regions. However, there is freedom to change the structure of the unit cells arbitrarily in the simulations of the edges. For instance, one may use the presented technique to simulate a periodic structure terminated by an edge taper \cite{munk2003finite,gustafsson2006rcs,johnson2002adiabatic}.

\bibliographystyle{IEEEtran}
\bibliography{main}

\begin{thebibliography}{10}
\providecommand{\url}[1]{#1}
\csname url@samestyle\endcsname
\providecommand{\newblock}{\relax}
\providecommand{\bibinfo}[2]{#2}
\providecommand{\BIBentrySTDinterwordspacing}{\spaceskip=0pt\relax}
\providecommand{\BIBentryALTinterwordstretchfactor}{4}
\providecommand{\BIBentryALTinterwordspacing}{\spaceskip=\fontdimen2\font plus
\BIBentryALTinterwordstretchfactor\fontdimen3\font minus
  \fontdimen4\font\relax}
\providecommand{\BIBforeignlanguage}[2]{{%
\expandafter\ifx\csname l@#1\endcsname\relax
\typeout{** WARNING: IEEEtran.bst: No hyphenation pattern has been}%
\typeout{** loaded for the language `#1'. Using the pattern for}%
\typeout{** the default language instead.}%
\else
\language=\csname l@#1\endcsname
\fi
#2}}
\providecommand{\BIBdecl}{\relax}
\BIBdecl

\bibitem{taflove2005computational}
A.~Taflove and S.~C. Hagness, \emph{{Computational Electrodynamics: The
  Finite-Difference Time-Domain Method}}.\hskip 1em plus 0.5em minus
  0.4em\relax Artech House, 2005.

\bibitem{kogon2020fdtd}
A.~J. Kogon and C.~D. Sarris, ``{FDTD Modeling of Periodic Structures: A
  Review},'' \emph{arXiv preprint arXiv:2007.05091}, 2020.

\bibitem{yang2007simple}
F.~Yang, J.~Chen, R.~Qiang, and A.~Elsherbeni, ``{A Simple and Efficient
  FDTD/PBC Algorithm for Scattering Analysis of Periodic Structures},''
  \emph{Radio Science}, vol.~42, no.~04, pp. 1--9, 2007.

\bibitem{kokkinos2006periodic}
T.~Kokkinos, C.~D. Sarris, and G.~V. Eleftheriades, ``{Periodic FDTD Analysis
  of Leaky-Wave Structures and Applications to the Analysis of
  Negative-Refractive-Index Leaky-Wave Antennas},'' \emph{IEEE Transactions on
  Microwave Theory and Techniques}, vol.~54, no.~4, pp. 1619--1630, 2006.

\bibitem{xu2007finite}
F.~Xu, K.~Wu, and W.~Hong, ``{Finite-Difference Time-Domain Modeling of
  Periodic Guided-Wave Structures and its Application to the Analysis of
  Substrate Integrated Nonradiative Dielectric Waveguide},'' \emph{IEEE
  Transactions on Microwave Theory and Techniques}, vol.~55, no.~12, pp.
  2502--2511, 2007.

\bibitem{chan1995order}
C.~Chan, Q.~Yu, and K.~Ho, ``{Order-N Spectral Method for Electromagnetic
  Waves},'' \emph{Physical Review B}, vol.~51, no.~23, p. 16635, 1995.

\bibitem{usoff1994edge}
J.~M. Usoff and B.~A. Munk, ``{Edge Effects of Truncated Periodic Surfaces of
  Thin Wire Elements},'' \emph{IEEE Transactions on Antennas and Propagation},
  vol.~42, no.~7, pp. 946--953, 1994.

\bibitem{ishimaru1985finite}
A.~Ishimaru, R.~Coe, G.~Miller, and W.~Geren, ``{Finite Periodic Structure
  Approach to Large Scanning Array Problems},'' \emph{IEEE Transactions on
  Antennas and Propagation}, vol.~33, no.~11, pp. 1213--1220, 1985.

\bibitem{ko1988scattering}
W.~L. Ko and R.~Mittra, ``{Scattering by a Truncated Periodic Array},''
  \emph{IEEE Transactions on Antennas and Propagation}, vol.~36, no.~4, pp.
  496--503, 1988.

\bibitem{denison1995decomposition}
D.~R. Denison and R.~W. Scharstein, ``{Decomposition of the Scattering by a
  Finite Linear Array into Periodic and Edge Components},'' \emph{Microwave and
  Optical Technology Letters}, vol.~9, no.~6, pp. 338--343, 1995.

\bibitem{wu1970analysis}
C.~Wu, ``{Analysis of Finite Parallel-Plate Waveguide Arrays},'' \emph{IEEE
  Transactions on Antennas and Propagation}, vol.~18, no.~3, pp. 328--334,
  1970.

\bibitem{capolino2000frequency}
F.~Capolino, M.~Albani, S.~Maci, and L.~B. Felsen, ``{Frequency-Domain Green's
  Function for a Planar Periodic Semi-Infinite Phased Array. I. Truncated
  Floquet Wave Formulation},'' \emph{IEEE Transactions on Antennas and
  Propagation}, vol.~48, no.~1, pp. 67--74, 2000.

\bibitem{holter2002some}
H.~Holter and H.~Steyskal, ``{Some Experiences from FDTD Analysis of Infinite
  and Finite Multi-Octave Phased Arrays},'' \emph{IEEE Transactions on Antennas
  and Propagation}, vol.~50, no.~12, pp. 1725--1731, 2002.

\bibitem{munk1979plane}
B.~Munk and G.~Burrell, ``{Plane-Wave Expansion for Arrays of Arbitrarily
  Oriented Piecewise Linear Elements and its Application in Determining the
  Impedance of a Single Linear Antenna in a Lossy Half-Space},'' \emph{IEEE
  Transactions on Antennas and Propagation}, vol.~27, no.~3, pp. 331--343,
  1979.

\bibitem{li2008efficient}
D.~Li and C.~D. Sarris, ``{Efficient Finite-Difference Time-Domain Modeling of
  Driven Periodic Structures and Related Microwave Circuit Applications},''
  \emph{IEEE Transactions on Microwave Theory and Techniques}, vol.~56, no.~8,
  pp. 1928--1937, 2008.

\bibitem{capolino2007comparison}
F.~Capolino, D.~R. Jackson, D.~R. Wilton, and L.~B. Felsen, ``{Comparison of
  Methods for Calculating the Field Excited by a Dipole Near a 2-D Periodic
  Material},'' \emph{IEEE Transactions on Antennas and Propagation}, vol.~55,
  no.~6, pp. 1644--1655, 2007.

\bibitem{memarian2012evanescent}
M.~Memarian and G.~V. Eleftheriades, ``{Evanescent-to-Propagating Wave
  Conversion in Sub-Wavelength Metal-Strip Gratings},'' \emph{IEEE Transactions
  on Microwave Theory and Techniques}, vol.~60, no.~12, pp. 3893--3907, 2012.

\bibitem{liu2007far}
Z.~Liu, S.~Durant, H.~Lee, Y.~Pikus, N.~Fang, Y.~Xiong, C.~Sun, and X.~Zhang,
  ``{Far-Field Optical Superlens},'' \emph{Nano Letters}, vol.~7, no.~2, pp.
  403--408, 2007.

\bibitem{mateus2004ultrabroadband}
C.~F. Mateus, M.~C. Huang, Y.~Deng, A.~R. Neureuther, and C.~J. Chang-Hasnain,
  ``{Ultrabroadband Mirror using Low-Index Cladded Subwavelength Grating},''
  \emph{IEEE Photonics Technology Letters}, vol.~16, no.~2, pp. 518--520, 2004.

\bibitem{min2008all}
C.~Min, P.~Wang, C.~Chen, Y.~Deng, Y.~Lu, H.~Ming, T.~Ning, Y.~Zhou, and
  G.~Yang, ``{All-Optical Switching in Subwavelength Metallic Grating Structure
  Containing Nonlinear Optical Materials},'' \emph{Optics Letters}, vol.~33,
  no.~8, pp. 869--871, 2008.

\bibitem{luo2002all}
C.~Luo, S.~G. Johnson, J.~Joannopoulos, and J.~Pendry, ``{All-Angle Negative
  Refraction without Negative Effective Index},'' \emph{Physical Review B},
  vol.~65, no.~20, p. 201104, 2002.

\bibitem{teixeira1999causality}
F.~Teixeira and W.~C. Chew, ``{On Causality and Dynamic Stability of Perfectly
  Matched Layers for FDTD Simulations},'' \emph{IEEE Transactions on Microwave
  Theory and Techniques}, vol.~47, no.~6, pp. 775--785, 1999.

\bibitem{holter2002size}
H.~Holter and H.~Steyskal, ``{On the Size Requirement for Finite Phased-Array
  Models},'' \emph{IEEE Transactions on Antennas and Propagation}, vol.~50,
  no.~6, pp. 836--840, 2002.

\bibitem{zhou2008size}
J.~Zhou, T.~Koschny, M.~Kafesaki, and C.~M. Soukoulis, ``{Size Dependence and
  Convergence of the Retrieval Parameters of Metamaterials},'' \emph{Photonics
  and Nanostructures-Fundamentals and Applications}, vol.~6, no.~1, pp.
  96--101, 2008.

\bibitem{parsons2011copper}
J.~Parsons and A.~Polman, ``{A Copper Negative Index Metamaterial in the
  Visible/Near-Infrared},'' \emph{Applied Physics Letters}, vol.~99, no.~16, p.
  161108, 2011.

\bibitem{munk2003finite}
B.~A. Munk, \emph{{Finite Antenna Arrays and FSS}}.\hskip 1em plus 0.5em minus
  0.4em\relax John Wiley \& Sons, 2003.

\bibitem{gustafsson2006rcs}
M.~Gustafsson, ``{RCS Reduction of Integrated Antenna Arrays with Resistive
  Sheets},'' \emph{Journal of Electromagnetic Waves and Applications}, vol.~20,
  no.~1, pp. 27--40, 2006.

\bibitem{johnson2002adiabatic}
S.~G. Johnson, P.~Bienstman, M.~Skorobogatiy, M.~Ibanescu, E.~Lidorikis, and
  J.~Joannopoulos, ``{Adiabatic Theorem and Continuous Coupled-Mode Theory for
  Efficient Taper Transitions in Photonic Crystals},'' \emph{Physical Review
  E}, vol.~66, no.~6, p. 066608, 2002.

\end{thebibliography}

\end{document}